\documentclass[english,onecolumn, draftcls]{IEEEtran}
\usepackage[T1]{fontenc}
\usepackage[latin9]{inputenc}
\usepackage{amsthm}
\usepackage{amsmath}
\usepackage{amssymb}
\usepackage{graphicx}
\usepackage{esint}

\makeatletter
\theoremstyle{plain}
\newtheorem{thm}{\protect\theoremname}
\theoremstyle{definition}
\newtheorem{defn}[thm]{\protect\definitionname}
\theoremstyle{plain}
\newtheorem{lem}[thm]{\protect\lemmaname}
\theoremstyle{remark}
\newtheorem{rem}[thm]{\protect\remarkname}

\renewcommand{\fnum@figure}{Fig.~\thefigure}

\makeatother

\usepackage{babel}
\providecommand{\definitionname}{Definition}
\providecommand{\lemmaname}{Lemma}
\providecommand{\remarkname}{Remark}
\providecommand{\theoremname}{Theorem}

\begin{document}

\title{Extracting Wyner's Common Information Using Polar Codes and Polar
Lattices}

\author{Jinwen Shi, Ling Liu, and Cong Ling \emph{Member, IEEE}}
\maketitle
\begin{abstract}
Explicit constructions of polar codes and polar lattices for both
lossless and lossy Gray-Wyner problems are studied. Polar codes are
employed to extract Wyner's common information of doubly symmetric
binary source; polar lattices are then extended to extract that of
a pair of Gaussian sources or multiple Gaussian sources. With regard
to the discrete sources, the entire best-known region of the lossless
Gray-Wyner problem are achieved by specifying the test channels to
construct polar codes without time-sharing. As a result, we are able
to give an interpretation that the Wyner's common information remains
the same to the lossy case when the distortion is small \cite{GeXulossyCI}.
Finally, the entire best-known lossy Gray-Wyner region for discrete
sources can also be achieved using polar codes. With regard to the
Gaussian sources, the best-known lossy Gray-Wyner region for bivariate
Gaussian sources with a specific covariance matrix \cite{GeXulossyCI}
can be achieved by using polar lattices. Moreover, we prove that extracting
Wyner's common information of a pair of Gaussian sources is equivalent
to implementing the lossy compression for a single Gaussian source,
which implies that the common information can be extracted by a polar
lattice for quantization. Furthermore, we extend this result to the
case of multiple Gaussian sources.
\end{abstract}

\let\thefootnote\relax\footnote{This work was presented in part at the IEEE International Conference
on Wireless Communications and Signal Processing 2016, Yangzhou, China,
October 2016. This work was supported in part by the Engineering and
Physical Sciences Research Council (EPSRC) and the China Scholarship
Council.

Jinwen Shi, Ling Liu, and Cong Ling are with Dept. of Electronic and
Electrical Engineering, Imperial College London, UK. (e-mails: \{jinwen.shi12,
l.liu12\}@imperial.ac.uk, cling@ieee.org).}

\section{Introduction }

\IEEEPARstart{T}{his} paper is concerned with extracting Wyner's
common information contained in a pair of correlated sources $(X,Y)$
using polar codes. There are different ways to characterize the amount
of common information in the literature. Apart from Shannon's mutual
information \cite{shannon2001mathematical} and G\'{a}cs-K\"{o}rner's
common information \cite{gacs1973common}, Wyner proposed an alternative
definition to quantify the common information of $(X,Y)$ with finite
alphabet \cite{WynerCI} as 
\[
C(X,Y)=\inf_{X-W-Y}I(X,Y;W),
\]
where the infimum is taken over all $W$, such that $X-W-Y$ forms
a Markov chain. 

G\'{a}cs-K\"{o}rner's common information has been found valuable in
applications for network securities and key generation \cite{Ahlswede1993Part1,Ahlswede1998Part2,Maurer1993Secret}.
However, G\'{a}cs-K\"{o}rner's common information is quite restrictive
in that it is non-zero only when the joint distribution of $\left(X,Y\right)$
satisfies certain requirements. Moreover, G\'{a}cs-K\"{o}rner's definition
is confined only for discrete random variables. Therefore, we investigate
how to the extract Wyner's common information of both discrete and
continuous random variables in this paper. Notice that both G\'{a}cs-K\"{o}rner's
and Wyner's common information are defined from theoretical viewpoints.
They are still important in many practical applications including
the performance limits in databases for correlated sources and in
minimum cost routing for networks \cite{Viswanatha2011Gray}.

\begin{figure}
\centering{}\includegraphics[width=9cm]{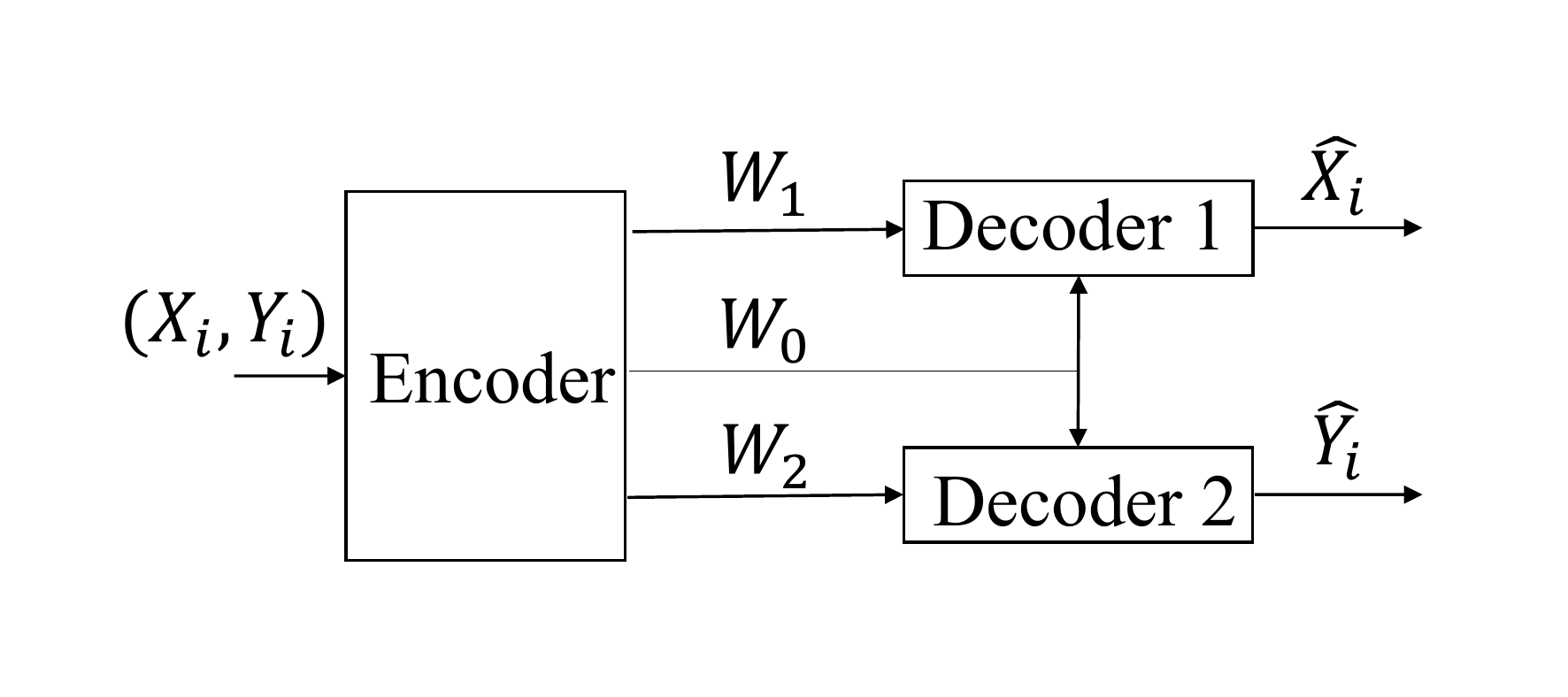}\protect\caption{Gray-Wyner source coding network.\label{fig:Gray-Wyner-source-coding-1}}
\end{figure}

\begin{figure}
\begin{centering}
\includegraphics[width=6cm]{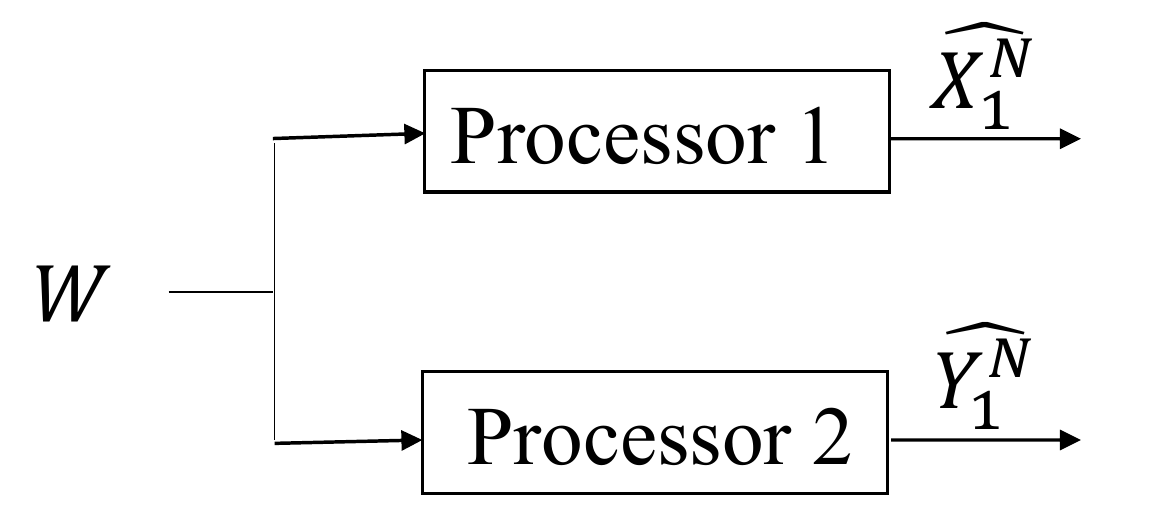}
\par\end{centering}

\protect\caption{The RV generators for Gray-Wyner source coding. \label{fig:The-RV-generators}}

\end{figure}

Wyner's definition originated from his earlier work on the Gray-Wyner
network \cite{gray1974source}, as depicted in Fig. \ref{fig:Gray-Wyner-source-coding-1},
which demonstrates Wyner's first approach \cite{WynerCI} to interpreting
$C(X,Y)$. This network model contains an encoder that observes a
pair of sequences $(X_{1}^{N},Y_{1}^{N})$ and outputs three messages
$W_{0}$, $W_{1}$ and $W_{2}$ with rate $R_{0},R_{1},R_{2}$, respectively.
Decoder 1 reconstructs $X_{1}^{N}$ by observing $(W_{0},W_{1})$
and decoder 2 reconstructs $Y_{1}^{N}$ from $(W_{0},W_{2})$. Wyner
also gave a second interpretation of the common information. In that
model, a common message $W$ is sent to two independent processors
as depicted in Fig. \ref{fig:The-RV-generators}. The processors generate
output sequences separately according to distributions $P_{X|W}(x|w)$
and $P_{Y|W}(y|w)$. The output sequences $\hat{X}_{1}^{N}$ and $\hat{Y}_{1}^{N}$
frame a joint probability 
\[
P_{\hat{X}_{1}^{N}\hat{Y}_{1}^{N}}(\hat{x}_{1}^{N},\hat{y}_{1}^{N})=\sum_{w\in\mathcal{W}}\frac{1}{|\mathcal{W}|}P_{X_{1}^{N}|W}(\hat{x}_{1}^{N}|w)P_{Y_{1}^{N}|W}(\hat{y}_{1}^{N}|w).
\]
Wyner showed that $C(X,Y)$ equals the minimum rate on the shared
message, on condition that the sum of rates equals the joint entropy
or that the joint distribution $P_{\hat{X}_{1}^{N}\hat{Y}_{1}^{N}}(\hat{x}_{1}^{N},\hat{y}_{1}^{N})$
is arbitrarily close to $P_{X_{1}^{N}Y_{1}^{N}}(x_{1}^{N},y_{1}^{N})$.

Wyner and G\'{a}cs-K\"{o}rner's work on common information can be
considered two different viewpoints of the lossless Gray-Wyner region.
Their works were then extended by \cite{viswanatha2014lossy} to
the lossy case, where the output sequences $(\hat{X}_{1}^{N},\hat{Y}_{1}^{N})$
have certain distortions. Moreover, a generalized lossy source coding
interpretation of Wyner's common information was given in \cite{GeXulossyCI}
for multiple dependent random variables with arbitrary number of alphabets. 

Wyner's common information of two Gaussian random variables was presented
in \cite{GeXulossyCI,viswanatha2014lossy}. A generalized formula
of Wyner's common information of jointly Gaussian vectors was deduced
in \cite{satpathy2015gaussian}. The dual problem was considered in
\cite{yang2014wyner}, where the common information of the outputs
of two additive Gaussian channels with a common input was computed.
For general continuous sources, the upper bound on the Wyner's common
information of multiple continuous random variables has been established
in terms of the dual total correlation in \cite{li2016distributed}.
In this paper, Wyner's common information of two or more Gaussian
sources presented in \cite{GeXulossyCI} will be extracted using polar
lattices. 

Polar codes \cite{arikan2009channel} have been widely studied due
to their achievability of Shannon bounds with low complexity. For
discrete sources, \cite{korada2009polar} provided constructions with
polar codes for lossless and lossy compression; \cite{PolarAsymOrig}
proposed a channel coding scheme for asymmetric settings using a concatenation
of two polar codes, and \cite{aspolarcodes} gave a solution to lossy
compression for nonuniform sources by a single polar code. For memoryless
Gaussian sources, \cite{PolarlatticeQZ} proposed a polar lattice
construction to achieve the rate-distortion bound.

The use of polar codes for the common information (i.e. point G in
Fig. \ref{fig:Performance-of-polar}) was recently proposed in \cite{GoelaCommonInfor},
which discussed polarization from the perspective of the maximal correlation
of two discrete sources. Furthermore, it proved that polar codes are
optimal to extract Wyner's common information of discrete sources. 

In this paper, we will investigate the entire best-known Gray-Wyner
region in \cite{WynerCI,gray1974source} for discrete sources. We
also give an interpretation to the results in \cite{GeXulossyCI}
that the common information defined for lossless Gray-Wyner coding
remains the same in the lossy case when the distortion is small. In
addition, an explicit construction based on polar codes and polar
lattices is given to achieve the entire lossy Gray-Wyner region \cite{GeXulossyCI}
for both doubly symmetric binary source (DSBS) and Gaussian sources. 

The main contributions of this paper are two-fold:
\begin{itemize}
\item The entire best-known lossless Gray-Wyner region in \cite{WynerCI,gray1974source}
is achieved by using polar codes. Moreover, based on the test channels
to construct polar codes, the entire region can be achieved without
time-sharing. In this case, the relations of the sub-regions of lossless
Gray-Wyner coding can be better understood. As a result, we are able
to give an interpretation of \cite{GeXulossyCI} that Wyner's common
information remains the same in the lossy case when the distortion
is small, from the perspective of source polarization.
\item An explicit construction based on polar codes is given to achieve
the lossy Gray-Wyner region \cite{GeXulossyCI} for a DSBS. In addition,
the lossy Gray-Wyner region \cite{GeXulossyCI} for two Gaussian sources
can be achieved by a construction of polar lattices. For both DSBS
and Gaussian sources, the lossy Gray-Wyner region not only contains
the case where lossy common information equals lossless common information,
but also the case where lossy common information equals the optimal
rate for a certain distortion pair of the source. Finally, the Wyner's
common information of multiple Gaussian random variables can also
be achieved by employing a polar lattice construction for Gaussian
random variables with a specific covariance matrix. 
\end{itemize}

The paper is organized as follows: Section II presents the background
of lossless and lossy compression using polar codes. The construction
of polar codes for the lossless Gray-Wyner network is investigated
in Section III. In Section IV, we construct polar codes for a DSBS
for the lossy Gray-Wyner network, and show simulation results for
different distortion regions. In Section V, we construct polar codes
for a pair of Gaussian variables for the lossy Gray-Wyner network;
then we extend the method to multiple Guassian sources. Finally, the
paper is concluded in Section VI.

\textit{Notations:} All random variables (RVs) are denoted by capital
letters. $P_X$  denotes the probability distribution of a RV $X$
taking value $x$ in a set $\mathcal{X}$. $X_{1}^{N}$ denotes a
vector $(X_{1},...,X_{N})$. For a set $\mathcal{I}$, $\mathcal{I}^c$
denotes its complement, and $|\mathcal{I}|$ denotes its cardinality.
$X_{\mathcal{I}}$ denotes the subvector $\{X_{i}\}_{i\in\mathcal{I}}$.
For an integer $N$, $[N]$ will be used to denote the set of all
integers from $1$ to $N$.  The information is measured in bits
and $h(\cdot)$ denotes the binary entropy function.

\section{Polar Codes for Source Coding}

\subsection{Polar Codes for Lossless Source Coding \label{sub:PCforLossless}\cite{korada2009polar}}

Let $X_{1}^{N}$ to be $N$ i.i.d. drawings of a RV $X$, where $X$
is a Bernoulli source with crossover probability $p$ (Ber($p$)).
$N=2^{n}$ for any integer $n\geq1$. The polarizing transformation
is defined by 
\[
U_{1}^{N}\triangleq X_{1}^{N}G_{N},
\]
where $G_{N}=G_{2}^{\otimes n}$ is the $n$-fold Kronecker product
of matrix 
\[
G_{2}=\begin{bmatrix}1 & 0\\
1 & 1
\end{bmatrix}.
\]
Fix $R>H(X)$ and let $\mathcal{F}$ denote the frozen set such that
$|\mathcal{F}|=\left\lceil NR\right\rceil $ and $H(U_{i}|U_{1}^{i-1})\geq H(U_{j}|U_{1}^{j-1})$
for all $i\in\mathcal{F}$, $j\notin\mathcal{F}$. The information
set is given by $\mathcal{I}=\mathcal{F}^{c}$. 

The Successive Cancellation (SC) encoder introduced in \cite{cronie2010lossless}
stores $u_{\mathcal{F}}$ and computes $\hat{u}_{\mathcal{I}}$, following
the encoding rules (\ref{eq:Lossy encode}) explained in Section \ref{sub:PCforLossyCompression}.
If $\hat{u}_{i}\neq u_{i}$ for $i\in\mathcal{I}$, an estimation
error occurs and the index $i$ needs to be announced to the decoder.
The set of error indices is denoted by $T$. The encoder outputs $(u_{\mathcal{F}},T)$.
The decoder puts $\hat{u}_{i}=u_{i}$ for $i\in\mathcal{F}$, then
estimates $u_{\mathcal{I}}$ using the same rule to the SC encoder
(\ref{eq:Lossy encode}). If $i\in T$, the decision is flipped. In
the end, the decoder outputs $\hat{x}_{1}^{N}=\hat{u}_{1}^{N}G_{N}$.
It has been shown in \cite{cronie2010lossless} that the error rate
tends to zero for any rate $R>H(X)$.

Since the entropy $H(U_{i}|U_{1}^{i-1})$ is complicated to analyze
when $N$ becomes very large, the Bhattacharyya parameter is often
used. For source coding with side information, assume $(X,Y)\in\{0,1\}\times\mathcal{Y}$
be a pair of RVs. The Bhattacharyya parameter \cite{arikan2010source}
is defined as 
\[
Z(X|Y)\triangleq2\sum_{y}P_{Y}(y)\sqrt{P_{X|Y}(0|y)P_{X|Y}(1|y)}.
\]
$Z(X|Y)$ and $H(X|Y)$ are related by \cite[Proposition 2]{arikan2010source},
which indicates that $H(X|Y)$ is near $0$ or $1$ if and only if
$Z(X|Y)$ is near $0$ or $1$, respectively. Hence, the parameters
$\{H(U_{i}|U_{1}^{i-1})\}_{1}^{N}$ and $\{Z(U_{i}|U_{1}^{i-1})\}_{1}^{N}$
polarize simultaneously. 

Furthermore, \cite{aspolarcodes,polarlatticeJ} show that the Bhattacharyya
parameter of an asymmetric channel can be equalized to the one in
the symmetric case. Therefore we can apply the known results in constructing
polar codes for symmetric channels \cite{Ido} to that for asymmetric
channels.

\subsection{Polar Codes for Lossy Source Coding \label{sub:PCforLossyCompression}\cite{korada2009polar,aspolarcodes}}

In this subsection, we discuss the lossy source coding for a nonuniform
source. We model the source as a sequence of i.i.d. realizations of
a RV $Y \in \mathcal{Y}$. $\mathcal{X}$ denotes the reconstruction
space. Denote the distortion function by $\ensuremath{\text{d}:\mathcal{Y}\times\mathcal{X}\rightarrow\mathbb{R}_{+}}$.
The rate-distortion function is given by 
\[
R(D)=\min_{\mathtt{\mathsf{E_{XY}}}[d(Y,X)]\leq D}I(X,Y).
\]

Similarly to lossless source coding, $U_{1}^{N}$ is defined as $U_{1}^{N}\triangleq X_{1}^{N}G_{N}$.
The frozen set $\mathcal{I}^{c}$ can be identified by 
\[
Z(U_{i}|U_{1}^{i-1},Y_{1}^{N})\geq1-2^{-N^{\beta}}\textrm{ or }Z(U_{i}|U_{1}^{i-1})\leq2^{-N^{\beta}}
\]
for all $i\in\mathcal{I}^{c}$. As a result, the information set satisfies
$|\mathcal{I}|=NR$, where $R$ is the encoding rate. \cite{aspolarcodes}
has shown that such an information set exists if $R>I(X,Y)=R(D)$,
$\beta<1/2$ and $N$ is sufficiently large. 

Once the indices of the information set and the frozen set are identified,
the encoder determines $u_{1}^{N}$ from a given source sequence $y_{1}^{N}$
by the rules
\begin{equation}
u_{i}=\begin{cases}
0, & \textrm{with probability }P_{U_{i}|U_{1}^{i-1},Y_{1}^{N}}(0|u_{1}^{i-1},y_{1}^{N})\\
1, & \textrm{with probability }P_{U_{i}|U_{1}^{i-1},Y_{1}^{N}}(1|u_{1}^{i-1},y_{1}^{N})
\end{cases}\label{eq:Lossy encode}
\end{equation}
for $i\in\mathcal{I}$ and 
\begin{equation}
u_{i}=\begin{cases}
\bar{u}_{i}, & \textrm{if }Z(U_{i}|U_{1}^{i-1},Y_{1}^{N})\geq1-2^{-N^{\beta}}\\
\arg\max_{u}P_{U_{i}|U_{1}^{i-1}}(u|u_{1}^{i-1}), & \textrm{if }Z(U_{i}|U_{1}^{i-1})\leq2^{-N^{\beta}}
\end{cases}\label{eq:Lossy_encoding_setting}
\end{equation}
for $i\in\mathcal{I}^{c}$. $\bar{u}_{i}$ is determined beforehand
uniformly from $\{0,1\}$, and shared between the encoder and the
decoder. Although the encoding scheme in \cite[ Theorem 4]{aspolarcodes}
is proved using a randomized map shared between the encoder and the
decoder, the alternative rule (\ref{eq:Lossy_encoding_setting}) in
our scheme has also been proposed in \cite{aspolarcodes} and further
proved in \cite{PolarlatticeQZ}. In the end, the encoder sends $u_{\mathcal{I}}$
to the decoder and the decoder outputs the reconstructed sequence
$x_{1}^{N}=u_{1}^{N}G_{N}$. 

Finally, it has been proved in \cite{aspolarcodes} that the average
distortion between the source and the reconstruction can be equivalent
to $D+O(2^{-N^{\beta'}})$ where $\beta'<\beta<1/2$. Notice that
the above construction is also applicable to symmetric sources, with
the index set $\{i:Z(U_{i}|U_{1}^{i-1})\leq2^{-N^{\beta}}\}=\textrm{Ø}$.
Additionally, the rule (\ref{eq:Lossy encode}) reduces to that of
lossless coding in the absence of $y_{1}^{N}$.

\section{Polar Codes for Lossless Gray-Wyner Coding}

\begin{figure}
\begin{centering}
\includegraphics[scale=0.6]{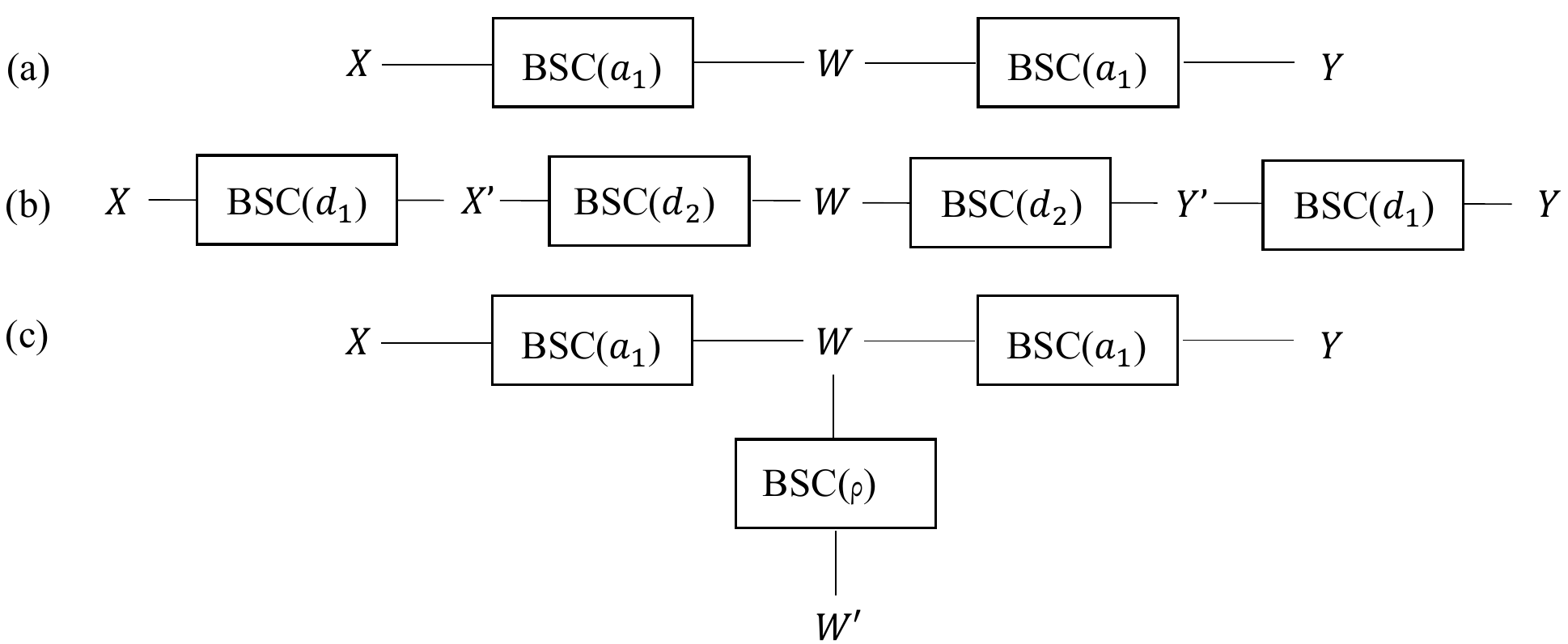}
\par\end{centering}

\protect\caption{Test channel for (a) Wyner's common information for DSBS source, where
$a_{0}=a_{1}*a_{1}$. (b) Line AG and lossy Gray-Wyner coding, where
$a_{1}=D_{1}*D_{2}$. (C) Line GB. ($a*b\triangleq a(1-b)+b(1-a)$).\label{fig:Test-channel} }
\end{figure}

In this section, we use the DSBS as an example to show that polar
codes are able to achieve the rate region of the Gray-Wyner network
in the lossless case. Consider the lossless coding model using Wyner's
first approach depicted in Fig. \ref{fig:Gray-Wyner-source-coding-1}.
Define the measurement of distortions as 
\[
d^{N}(y_{1}^{N},x_{1}^{N})\triangleq\sum_{i=1}^{N}d(y_{i},x_{i}),
\]
where $d(y,x)$ denotes Hamming distance for discrete RVs or Euclidean
distance for continuous RVs. Following that, we give a formal definition
to this model.
\begin{defn}
\label{def:Gray-Wyner network}An $(N,M_{0},M_{1},M_{2})$ code for
the lossless Gray-Wyner coding depicted in Fig. \ref{fig:Gray-Wyner-source-coding-1}
is defined as follows: 

An encoder is a mapping 
\[
f_{E}:\mathcal{X}^{N}\times\mathcal{Y}^{N}\rightarrow I_{M_{0}}\times I_{M_{1}}\times I_{M_{2}},
\]
where $I_{M_{i}}=\{0,1,2,\ldots,M_{i}-1\}$ for $i=0,1,2$. 

A decoder is a pair of mappings 
\[
\begin{aligned} & f_{D}^{(X)}:I_{M_{0}}\times I_{M_{1}}\rightarrow\mathcal{X}^{N}\\
 & f_{D}^{(Y)}:I_{M_{0}}\times I_{M_{2}}\rightarrow\mathcal{Y}^{N}.
\end{aligned}
\]

Let $f_{E}(X_{1}^{N},Y_{1}^{N})=(W_{0},W_{1},W_{2})$, then 
\[
\begin{aligned} & \hat{X}_{1}^{N}=f_{D}^{(X)}(W_{0},W_{1})\\
 & \hat{Y}_{1}^{N}=f_{D}^{(Y)}(W_{0},W_{2}).
\end{aligned}
\]
The average distortion between the inputs and outputs are $(\triangle_{X},\triangle_{Y})$,
where 
\[
\begin{aligned} & \triangle_{X}=\frac{1}{N}\mathsf{E}d^{N}(X_{1}^{N},\hat{X}_{1}^{N})\\
 & \triangle_{Y}=\frac{1}{N}\mathsf{E}d^{N}(Y_{1}^{N},\hat{Y}_{1}^{N}).
\end{aligned}
\]

\end{defn}
The achievable rate region of an $(N,M_{0},M_{1},M_{2})$ code is
defined as follows.
\begin{defn}
\label{def:lossless DSBS}For lossless coding, a triple $(R_{0},R_{1},R_{2})$
is said to be $\mathit{achievable}$ if there exists an $(N,M_{0},M_{1},M_{2})$
code with $M_{i}\leq2^{N(R_{i}+\epsilon)}$, $i=0,1,2$ and $\bigtriangleup=\max(\triangle_{X},\triangle_{Y})\leq\epsilon$,
for arbitrary $\epsilon>0$ and sufficiently large $N$. Denote $\mathcal{R}$
as the set of achievable rate. \end{defn}
\begin{thm}
\label{thm:(R0R1R2)}(\cite[Theorem 2]{gray1974source}) If $(R_{0}$,$R_{1}$,$R_{2})\in\mathcal{R}$,
then 
\[
\begin{cases}
R_{0}+R_{1}+R_{2}\geq H(X,Y)\\
R_{0}+R_{1}\geq H(X)\\
R_{0}+R_{2}\geq H(Y).
\end{cases}
\]
 
\end{thm}
Let us consider a DSBS $(X,Y)$, where $\mathcal{X}=\mathcal{Y}=\{0,1\}$
and $Y=X\oplus Z$, $Z\sim\textrm{Ber}(a_{0})$. In this case, $H(X,Y)=1+h(a_{0})$,
$H(X)=H(Y)=1$. It has been shown in \cite{WynerCI} that the common
information of DSBS is $C(X,Y)=1+h(a_{0})-2h(a_{1})$, where $a_{1}=\frac{1}{2}-\frac{1}{2}(1-2a_{0})^{1/2}$.
This model can be considered a cascade of two Binary Symmetric Channels
(BSCs) with the same crossover probability $a_{1}$. The cascaded
channel is equivalent to a single BSC($a_{0}$). It was shown in \cite{WynerCI}
that the common information $C(X,Y)$ can be achieved when $W$ is
an intermediate RV as depicted in Fig. \ref{fig:Test-channel} (a).

\begin{figure}
\centering{}\includegraphics[scale=0.6]{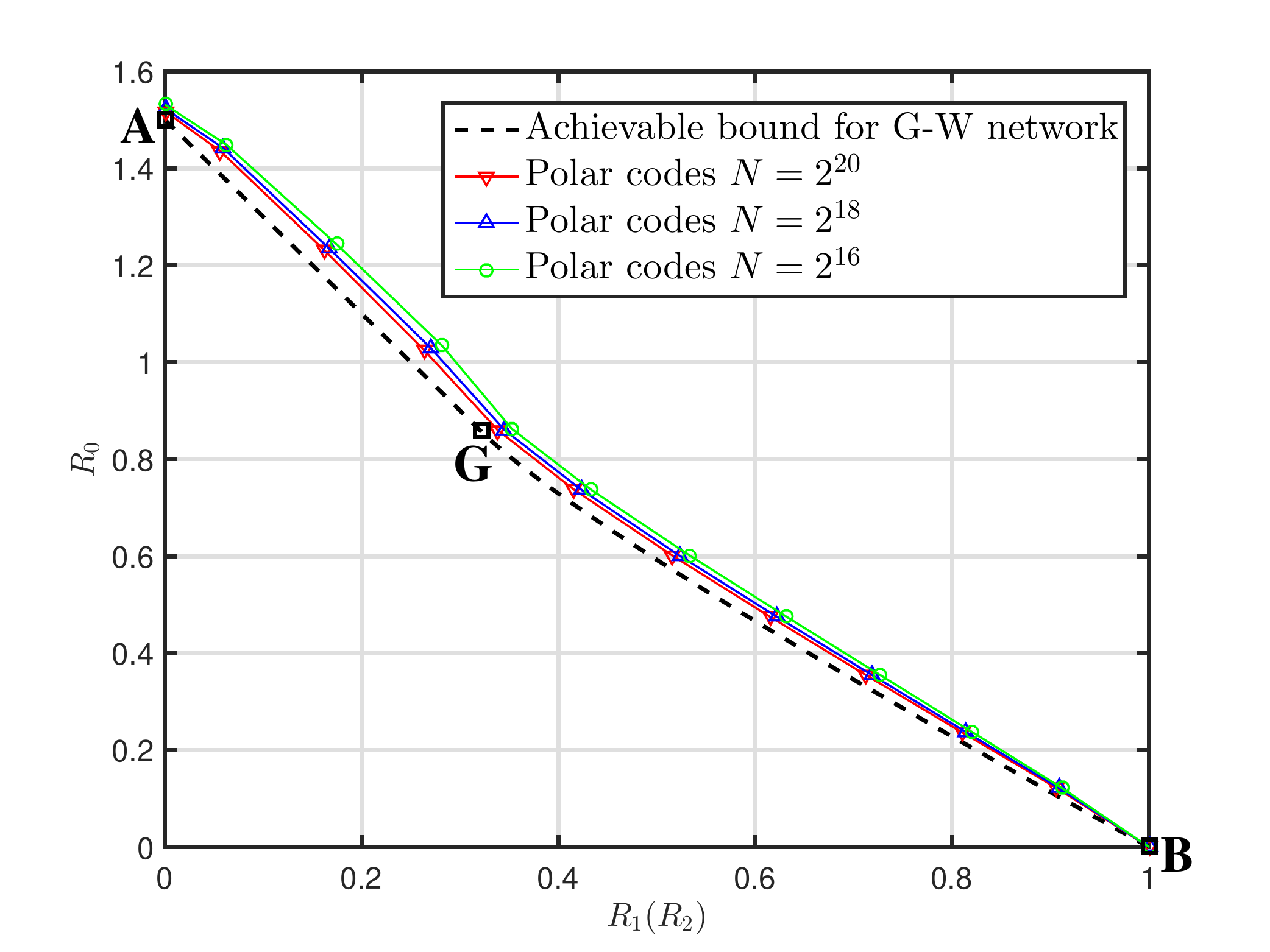}\protect\caption{Simulation performance of polar codes for Gray-Wyner lossless coding
when $a_{0}=0.11$ and $N=2^{16},2^{18},2^{20}$. The performance
loss between AG is due to the nonuniform source coding.\label{fig:Performance-of-polar}}
\end{figure}

\subsection{Polar codes for Pangloss bound\label{sub:PanglossBound} }

Theorem \ref{thm:(R0R1R2)} indicates a fact that the Gray-Wyner network
defined in Definition \ref{def:lossless DSBS} cannot perform better
than the situation where the receivers can collaborate. This situation
is referred to as the Pangloss plane \cite{gray1974source}, where
the triple $(R_{0},R_{1},R_{2})$ satisfies 
\[
\sum_{i=0}^{2}R_{i}=H(X,Y).
\]
The dashed line AG in Fig. \ref{fig:Performance-of-polar} is the
Pangloss bound for the lossless Gray-Wyner problem. Point A refers
to the case where only the common branch is used. Therefore the problem
is the same to the joint compression for source $(X,Y)$ when $(R_{0},R_{1},R_{2})=(H(X,Y),0,0)$.
Point G refers to the case where $R_{0}$ is the smallest rate that
achieves lossless compression for source $(X,Y)$ when the total rate
equals $H(X,Y)$. In other words, $R_{0}$ achieves Wyner's common
information $C(X,Y)$ at point G and $(R_{0},R_{1},R_{2})=(1+h(a_{0})-2h(a_{1}),h(a_{1}),h(a_{1}))$.
Next we demonstrate how to achieve the Pangloss bound using polar
codes without time-sharing.
\begin{itemize}
\item Point A: 
\end{itemize}
This point can be trivially achieved, since 
\[
H(X,Y)=H(X)+H(Y|X)=H(X)+H(Z).
\]
In this case, $X$ does not need to be compressed. Encoder sends $x_{1}^{N}$
and compresses $z_{1}^{N}=x_{1}^{N}\oplus y_{1}^{N}$ as introduced
in Section \ref{sub:PCforLossless}. Decoders 1 and 2 reconstruct
$\hat{x}_{1}^{N}$ and $\hat{z}_{1}^{N}$ with error probability tending
to zero when $N$ is sufficiently large. Then $\hat{y}_{1}^{N}=\hat{x}_{1}^{N}\oplus\hat{z}_{1}^{N}$.
Due to symmetry, the role of source $X$ and $Y$ can be exchanged.
\begin{itemize}
\item Point G: 
\end{itemize}
Firstly we extract the common information from source $(X_{1}^{N},Y_{1}^{N})$.
Since 
\[
R_{0}\geq I(W;XY)=1+h(a_{0})-2h(a_{1}),
\]
the encoder applies lossy compression to the joint sources $(x_{1}^{N},y_{1}^{N})$
as introduced in Section \ref{sub:PCforLossyCompression} with reconstruction
$w_{1}^{N}$. Differently from traditional lossy compression with
a single source, the test channel for the joint sources is 
\[
P_{XY|W}(xy|w)=P_{X|W}(x|w)P_{Y|W}(y|w).
\]

In this way, the average distortion between $w_{1}^{N}$ and $x_{1}^{N}$
or $w_{1}^{N}$ and $y_{1}^{N}$ will tend to $a_{1}$ simultaneously
as $N$ increases. Hence the lossy-compressed sequence $u_{\mathcal{I}}$
is the common message and is sent to both decoders, where $|\mathcal{I}|\geq(1+h(a_{0})-2h(a_{1}))N$.
Since 
\[
H(X|W)=H(Y|W)=h(a_{1}),
\]
we apply lossless compression to source $x_{1}^{N}$ together with
$w_{1}^{N}$ as side information and send the lossless compressed
sequence privately to Decoder 1. Symmetrically, source $Y$ is operated
in the same way. 

At the decoder side, both decoders reconstruct $w_{1}^{N}$ from $u_{\mathcal{I}}$
by the lossy decoding rule introduced in Section \ref{sub:PCforLossyCompression}.
Decoder 1 receives the compressed sequence from its private branch
and derives a reconstructed sequence $\hat{z}_{1}^{N}$. Then the
source can be reconstructed as $\hat{x}_{1}^{N}=\hat{z}_{1}^{N}\oplus w_{1}^{N}$.
Decoder 2 operates in the same way. Therefore, $X$ and $Y$ can be
reconstructed with error rate tending to zeros when $N$ is sufficiently
large. Notice that a similar method was also given in \cite{GoelaCommonInfor}.
\begin{itemize}
\item Points on dashed line AG: 
\end{itemize}
On this line, the common branch carries more information than point
G. In order to show how much additional amount of information to be
sent over the common branch, we keep the relation that $X-W-Y$ is
a Markov chain. Now, if we move the intermediate RV $W$ closer to
sources $(X,Y)$, there will be two new RVs $(X',Y')$ between the
source and common RV $W$ correspondingly. Hence we assume that the
test channel is a BSC($d_{1}$) between $X$ ($Y$) and $X'$ ($Y'$),
and a BSC($d_{2}$) between $W$ and $(X'$,$Y')$ where $0\leq d_{1},d_{2}\leq a_{1}$.
This test channel is depicted in Fig. \ref{fig:Test-channel} (b).
Therefore, 
\[
X-X'-W-Y'-Y
\]
forms a Markov chain. 

In this case, the rate of the common branch is 
\[
\begin{aligned}R_{0} & \geq I(X'Y';XY)\\
 & =I(X'Y'W;XY)\\
 & =I(XY;W)+I(X';X|W)+I(Y';Y|W)\\
 & =1-2h(d_{1})+h(a_{0}).
\end{aligned}
\]
Instead of extracting the common RV $W$ over the common branch, decoders
can reconstruct $(X',Y')$ and retrieve more information from the
common branch. This is because $(X',Y')$ is closer to $(X,Y)$ than
$W$ in the Markov chain. Hence, both sources can be losslessly reconstructed
by applying lossless compression to sources $(X,Y)$ with side information
$(X',Y')$. Therefore the rate of the private branches is 
\[
R_{1}=R_{2}\geq H(X|X')=H(Y|Y')=h(d_{1}).
\]

Next we show a construction using polar codes that achieves the rate
bound. Similarly to point G, we can firstly retrieve the common message
$w_{1}^{N}$ by applying lossy compression to joint source sequences
$(x_{1}^{N},y_{1}^{N})$. The compressed rate approaches $1+h(a_{0})-2h(a_{1})$.
Due to that $Z_{1}=X\oplus W$ is a Ber($a_{1}$) source, the encoder
applies lossy compression to the nonuniform source realizations $x_{1}^{N}\oplus w_{1}^{N}$
with distortion $d_{1}$ as introduced in Section \ref{sub:PCforLossyCompression}.
We operate the same to source $Y$. Therefore, the additional rate
approaches $2(h(a_{1})-h(d_{1}))$ when $N$ is sufficiently large.
For the private branches, the encoder firstly reconstructs $X'$ and
$Y'$. Then the encoder applies lossless compression to source $X\oplus X'$
and $Y\oplus Y'$ with rate approaching $h(d_{1})$. It is known that
polar codes are optimal for lossy and lossless compression \cite{aspolarcodes,arikan2010source}.
Therefore, the average distortion of sources $(X_{1}^{N},Y_{1}^{N})$
approaches zero when $N$ is sufficiently large.

\subsection{Polar codes for $R_{0}<C(X,Y)$ (Curve GB)}

In this part, we show how to achieve the dashed curve GB in Fig. \ref{fig:Performance-of-polar}
using polar codes. From Theorem \ref{thm:(R0R1R2)}, the lower boundary
of $\mathcal{R}$ should lie above the lines 
\[
\begin{cases}
R_{0}+2R_{1}=1+h(a_{0})\\
R_{0}+R_{1}=1.
\end{cases}
\]
In the preceding subsection, we have shown constructions to achieve
this lower boundary of $\mathcal{R}$, called Pangloss bound, when
$R_{0}\geq C(X,Y)$ (e.g. dashed line AG in Fig. \ref{fig:Performance-of-polar}).
However the lower boundary of $\mathcal{R}$ remains unknown when
$R_{0}<C(X,Y)$. To the best of our knowledge, the tightest lower
boundary was given in \cite{gray1974source} for the case as follows:

Consider a degradation applied on the common RV $W$, where $W$ is
considered the input to BSC($\rho$) with output $W'$ as shown in
Fig. \ref{fig:Test-channel} (C). As a result, 
\[
(X,Y)-W-W'
\]
forms a Markov chain and $a_{1}\leq\beta=a_{1}*\rho\leq1/2$. 

From this Markov chain, the transition probability reads

\begin{align}
P_{XY|W'}(x,y|w') & =\sum_{w\in\mathcal{W}}P_{W|W'}(w|w')P_{X|W}(x|w)P_{Y|W}(y|w).\label{eq:Transition probability}
\end{align}
Then the triple $(R_{0},R_{1},R_{2})\in\mathcal{R}$ satisfies 
\begin{align}
\begin{cases}
R_{0}=I(W';XY)=(1-a_{0})(1-h(\frac{\beta-0.5a_{0}}{1-a_{0}}))\\
R_{1}=R_{2}=H(X|W')=H(Y|W')=h(\beta).
\end{cases}\label{eq:TriplesForSamllR_0}
\end{align}

As $\beta\in[a_{1},\frac{1}{2}]$, the family of rate triples can
generate the dashed curve GB in Fig. \ref{fig:Performance-of-polar}.
In fact, achieving the triple in (\ref{eq:TriplesForSamllR_0}) is
quite similar to achieving the rate bound for point G using polar
codes. Firstly we apply lossy compression to joint sources $(X_{1}^{N},Y_{1}^{N})$
with distortion $\beta$ and derive reconstruction $W'$. The test
channel used in lossy compression is specified in (\ref{eq:Transition probability}),
which is the major difference from the construction of point G. Next,
send the compressed sequence over the common branch. After that, we
apply lossless compression to source $X$ and $Y$ with $W'$ as side
information, and send the compressed sequences through private branches
to decoder 1 and 2 accordingly. Alternatively we can derive the common
RV $W$ by lossy compression of $(X,Y)$. Afterwards, we apply symmetric
lossy compression to $W$ with distortion $\rho$ to obtain $W'$.
Finally, it is trivial to achieve point $B$ when $R_{1}=R_{2}=1$
and $R_{0}=0$. 

Together with the result from \ref{sub:PanglossBound}, all points
from point A to B along the dashed line in Fig. \ref{fig:Performance-of-polar}
can be achieved by polar codes. Moreover, the above models can be
extended to achieving the lower boundary of more general binary-correlated
sources mentioned in \cite[Theorem 3]{witsenhausen1976values}.

\section{Lossy Gray-Wyner Coding for a DSBS}

In this section, we show how to achieve the lossy common information
$C\left(\triangle_{1},\triangle_{2}\right)$, which is defined as
the smallest common rate $R_{0}$, such that the total rate meets
the rate-distortion bound. Based on Definition \ref{def:Gray-Wyner network},
we define the lossy Gray-Wyner coding as follows: 

The rate-distortion function for source $(X,Y)$ is
\[
R_{XY}\left(\triangle_{1},\triangle_{2}\right)=\min_{P_{X'Y'|XY}(x'y'|xy):\mathsf{E}d(X',X)\leq\triangle_{1},\mathsf{E}d(Y',Y)\leq\triangle_{2}}I(X',Y';X,Y),
\]
where the minimum is taken over all test channels $P_{X'Y'|XY}(x'y'|xy)$
such that $\mathsf{E}d(X',X)\leq\triangle_{1}$ and $\mathsf{E}d(Y',Y)\leq\triangle_{2}$. 
\begin{defn}
For any $\triangle_{1},\triangle_{2}\geq0$, a number $R_{0}$ is
said to be $\left(\triangle_{1},\triangle_{2}\right)\mathit{-achivable}$
if for any $\varepsilon>0$ we can find a sufficiently large $N$
such that there exists a $(N,M_{0},M_{1},M_{2})$ code with 
\[
\begin{aligned}M_{0} & \leq2^{NR_{0}},\\
\sum_{i=0}^{2}\frac{1}{N}\log M_{i} & \leq R_{XY}\left(\triangle_{1},\triangle_{2}\right)+\varepsilon,\\
\triangle_{X} & \leq\triangle_{1}+\varepsilon,\\
\triangle_{Y} & \leq\triangle_{2}+\varepsilon.
\end{aligned}
\]
Then $C\left(\triangle_{1},\triangle_{2}\right)$ is defined as the
infimum of all $R_{0}$ that is $\left(\triangle_{1},\triangle_{2}\right)$-achievable. 
\end{defn}
To avoid ambiguity, we refer $C\left(\triangle_{1},\triangle_{2}\right)$
to lossy common information and $C\left(X,Y\right)$ to common information
or Wyner's common information in this section.

The authors  gave a characterization of $C\left(\triangle_{1},\triangle_{2}\right)$
in \cite{viswanatha2014lossy} as follows: 
\begin{defn}
Given a pair of joint sources $\left(X,Y\right)\sim P_{XY}(x,y)$,
for any $\triangle_{1},\triangle_{2}\geq0$, the lossy common information
reads 
\[
C\left(\triangle_{1},\triangle_{2}\right)=\inf I\left(X,Y;W\right),
\]
where the infimum is taken over all joint distributions for $X$,
$Y$, $X'$, $Y'$, $W$ such that 
\begin{equation}
\begin{aligned}X'-W-Y',\\
\left(X,Y\right)-\left(X',Y'\right)-W,
\end{aligned}
\label{eq:Character_lossyCI}
\end{equation}
and $\left(X',Y'\right)$ achieves $R_{XY}\left(\triangle_{1},\triangle_{2}\right)$. 
\end{defn}
Notice that this is a more generalized characterization, comparing
with that for lossless Gray-Wyner network where $X$ and $Y$ are
independent given $W$. 

Following the above definition, we consider the same DSBS $\left(X,Y\right)$
to the previous section. The joint distribution of $\left(X,Y\right)$
is given by 
\[
P_{X,Y}\left(x,y\right)=\begin{cases}
\frac{1}{2}\left(1-a_{0}\right), & \textrm{if}\textrm{ }x=y\\
\frac{1}{2}a_{0}, & \textrm{otherwise},
\end{cases}
\]
where $a_{0}\in\left[0,\frac{1}{2}\right]$. Let $a_{0}=2a_{1}\left(1-a_{1}\right)$
and $a_{1}\in\left[0,\frac{1}{2}\right]$.

\begin{figure}
\begin{centering}
\includegraphics[scale=0.7]{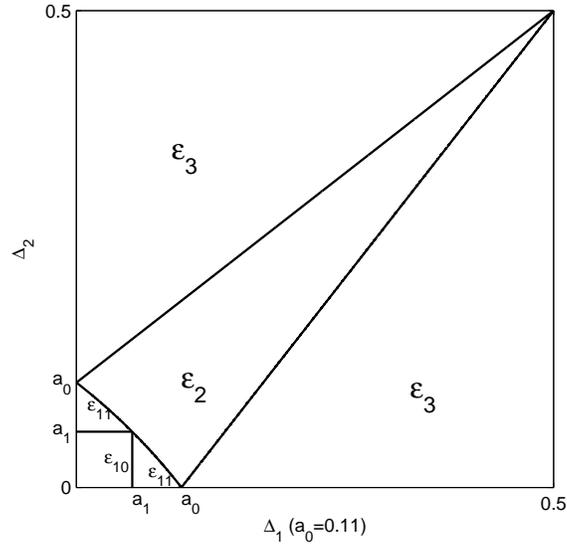}
\par\end{centering}

\protect\caption{The distortion regions for DSBS when $a_{0}=0.11$. \label{fig:The-distortion-regions-DSBS}}

\end{figure}

The lossy common information for $\left(X,Y\right)$ has been given
by \cite{GeXulossyCI} 
\[
C\left(\triangle_{1},\triangle_{2}\right)=\begin{cases}
C\left(X,Y\right) & \left(\triangle_{1},\triangle_{2}\right)\in\varepsilon_{10},\\
R_{XY}\left(\triangle_{1},\triangle_{2}\right) & \left(\triangle_{1},\triangle_{2}\right)\in\varepsilon_{2}\cup\varepsilon_{3}\\
0 & \left(\triangle_{1},\triangle_{2}\right)\geq\left(\frac{1}{2},\frac{1}{2}\right),
\end{cases},
\]

\[
C\left(X,Y\right)\leq C\left(\triangle_{1},\triangle_{2}\right)\leq R_{XY}\left(\triangle_{1},\triangle_{2}\right),\textrm{ }\textrm{ }\textrm{ }\left(\triangle_{1},\triangle_{2}\right)\in\varepsilon_{11},
\]
where
\begin{align*}
\varepsilon_{10} & =\left\{ \left(\triangle_{1},\triangle_{2}\right):0\leq\triangle_{i}\leq a_{1},\textrm{ }i=1,2\right\} ,\\
\varepsilon_{11} & =\varepsilon_{10}^{c}\cap\left\{ \left(\triangle_{1},\triangle_{2}\right):\triangle_{1}+\triangle_{2}-2\triangle_{1}\triangle_{2}\leq a_{0}\right\} ,\\
\varepsilon_{2} & =\varepsilon_{10}^{c}\cap\varepsilon_{11}^{c}\cap\left\{ \left(\triangle_{1},\triangle_{2}\right):\max\left\{ \frac{\triangle_{1}-\triangle_{2}}{1-2\triangle_{2}},\frac{\triangle_{2}-\triangle_{1}}{1-2\triangle_{1}}\right\} \leq a_{0}\right\} ,\\
\varepsilon_{3} & =\varepsilon_{10}^{c}\cap\varepsilon_{11}^{c}\cap\varepsilon_{2}^{c}\cap\left\{ \left(\triangle_{1},\triangle_{2}\right):\triangle_{i}\leq\frac{1}{2},\textrm{ }i=1,2\right\} .
\end{align*}
The relations among these four regions are depicted in Fig. \ref{fig:The-distortion-regions-DSBS}.
The joint rate distortion function of the DSBS $\left(X,Y\right)$
is given by \cite{nayak2010successive} 
\[
\begin{aligned} & R_{XY}\left(\triangle_{1},\triangle_{2}\right)=\begin{cases}
1+h\left(a_{0}\right)-h\left(\triangle_{1}\right)-h\left(\triangle_{2}\right), & \left(\triangle_{1},\triangle_{2}\right)\in\varepsilon_{1},\\
1-\left(1-a_{0}\right)h\left(\frac{\triangle_{1}+\triangle_{2}-a_{0}}{2\left(1-a_{0}\right)}\right)-a_{0}h\left(\frac{\triangle_{1}-\triangle_{2}+a_{0}}{2a_{0}}\right), & \left(\triangle_{1},\triangle_{2}\right)\in\varepsilon_{2},\\
1-h\left(\min\left\{ \triangle_{1},\triangle_{2}\right\} \right), & \left(\triangle_{1},\triangle_{2}\right)\in\varepsilon_{3},
\end{cases}\end{aligned}
\]
where $\varepsilon_{1}=\varepsilon_{10}\cup\varepsilon_{11}$. 

First, we provide an operational meaning of the region $\left(\triangle_{1},\triangle_{2}\right)\in\varepsilon_{10}$,
where $C\left(\triangle_{1},\triangle_{2}\right)=C(X,Y)$. 

The relation between lossy and lossless Gray-Wyner coding is not difficult
to find if we recall the construction of polar codes for the line
AG in Section \ref{sub:PanglossBound}. We apply the same test channel
where $X-X'-W-Y'-Y$ forms a Markov chain depicted by Fig. \ref{fig:Test-channel}
(b). The rate of the shared branch in lossless Gray-Wyner coding reads
\[
I(X'Y';XY)=I(XY;W)+I(X';X|W)+I(Y';Y|W).
\]
Thus we shall send all of the three sub-sequences over the shared
branch. 

Notice that if we consider $(X',Y')$ the output of the decoders,
the above rate is sufficient for lossy Gray-Wyner coding. In the lossy
case, we only require to recover the sources $(X,Y)$ with distortions
$\left(\triangle_{1},\triangle_{2}\right)$. Therefore, we consider
the intermediate RVs $(X',Y')$ as the reconstruction RVs.

Similar to the case in Section \ref{sub:PanglossBound}, we only consider
the plane in the $\left(R_{0},R_{1},R_{2}\right)$ space where $R_{1}=R_{2}$
and $\triangle_{1}=\triangle_{2}=\triangle$. The encoder applies
the same lossy compression as that for line AG to extract $W$, and
sends it on the shared branch with rate $R_{0}\geq I(XY;W)$. Next,
the encoder applies lossy compression to the nonuniform source $X\oplus W$
with distortion $\triangle$ where $a_{1}=d_{2}*\triangle$. To achieve
$\triangle_{1}=\triangle_{2}=\triangle$, the additional rate we should
send is $I(X';X|W)$ and $I(Y';Y|W)$ over either private branches
or the shared branch. Then the distortion between $X(Y)$ and $X'(Y')$
tends to $\triangle$ when $N$ is sufficiently large. As a result,
the total rate 
\[
\begin{aligned}\sum_{i=0}^{2}R_{i}\rightarrow I(X'Y';XY) & =1+h(a_{0})-2h(\triangle)=R_{XY}(\triangle,\triangle),\end{aligned}
\]
where $0\leq\triangle\leq a_{1}$. This indicates that the lossy Pangloss
bound \cite{gray1974source} can be achieved as long as $0\leq\triangle\leq a_{1}$
and $C\left(\triangle,\triangle\right)=C\left(X,Y\right)$.

For $\left(\triangle_{1},\triangle_{2}\right)\in\varepsilon_{2}\cup\varepsilon_{3}$
, the lossy common information exactly equals the optimal rate for
a certain distortion pair for the joint DSBS. This means that all
the messages should be sent over the shared branch to achieve the
desired distortion $\left(\triangle_{1},\triangle_{2}\right)\in\varepsilon_{2}\cup\varepsilon_{3}$. 

We then construct polar codes to extract the lossy common information
for $\left(\triangle_{1},\triangle_{2}\right)\in\varepsilon_{2}$.
Fortunately the backward test channel that achieves $R_{XY}\left(\triangle_{1},\triangle_{2}\right)$
has been given by \cite{nayak2010successive} 
\[
\begin{cases}
X=X'+Z_{1}\\
Y=Y'+Z_{2},
\end{cases}
\]
where $\left[X',Y'\right]^{\mathrm{T}}$ and $\left[Z_{1},Z_{2}\right]^{\mathrm{T}}$
are vectors contained binary RVs ($\cdot^{\mathrm{T}}$ stands for
matrix transpose), where the two vectors are independent of each other.
Additionally, the probability mass function are given by 
\[
P_{X',Y'}(x',y')=\begin{bmatrix}\frac{1}{2} & 0\\
0 & \frac{1}{2}
\end{bmatrix},\textrm{ }\textrm{ }\textrm{ }
\]

\[
P_{Z_{1},Z_{2}}(z_{1},z_{2})=\frac{1}{2}\begin{bmatrix}2-a_{0}-\triangle_{1}-\triangle_{2} & \triangle_{2}-\triangle_{1}+a_{0}\\
\triangle_{1}-\triangle_{2}+a_{0} & \triangle_{1}+\triangle_{2}-a_{0}
\end{bmatrix}.
\]

From the joint probability $P_{X',Y'}(x',y')$, it is trivial that
the condition for $\left(X',Y'\right)$ to achieve $R_{XY}\left(\triangle_{1},\triangle_{2}\right)$
is 
\[
X'=Y'.
\]
For simplicity, we assign a binary RV $W$ such that $W=X'=Y'$.

Similar to the code construction of Curve GB, we design polar codes
for performing the lossy compression that generates the reconstruction
$W$ on the joint sources $\left(X,Y\right)$. The transition probability
of the test channel is given by

\[
\begin{cases}
\begin{aligned}P_{XY|W}(0,0|1) & =P_{XY|W}(1,1|0)=P_{Z_{1}Z_{2}}(1,1)=\frac{1}{2}\left(\triangle_{1}+\triangle_{2}-a_{0}\right),\end{aligned}
\\
\begin{aligned}P_{XY|W}(0,1|1) & =P_{XY|W}(1,0|0)=P_{Z_{1}Z_{2}}(1,0)=\frac{1}{2}\left(\triangle_{1}-\triangle_{2}+a_{0}\right),\end{aligned}
\\
\begin{aligned}P_{XY|W}(1,0|1) & =P_{XY|W}(0,1|0)=P_{Z_{1}Z_{2}}(0,1)=\frac{1}{2}\left(\triangle_{2}-\triangle_{1}+a_{0}\right),\end{aligned}
\\
\begin{aligned}P_{XY|W}(1,1|1) & =P_{XY|W}(0,0|0)=P_{Z_{1}Z_{2}}(0,0)=\frac{1}{2}\left(2-a_{0}-\triangle_{1}-\triangle_{2}\right).\end{aligned}
\end{cases}
\]

\begin{figure}
\centering{}\includegraphics[scale=0.9]{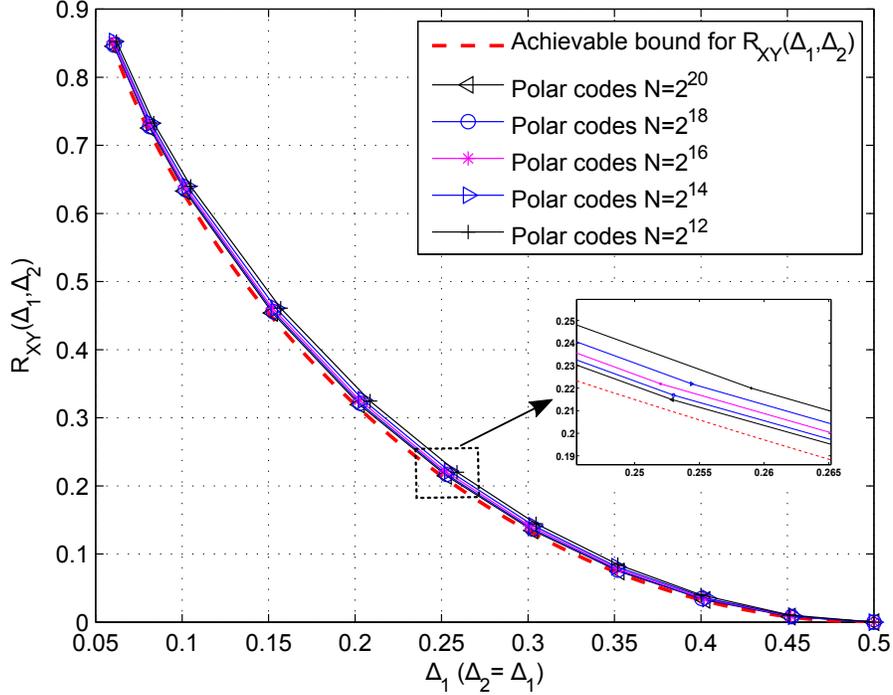}\protect\caption{Simulation performance for polar codes for Gray-Wyner lossy coding
when $\left(\triangle_{1},\triangle_{2}\right)\in\varepsilon_{2}$
, $\triangle_{1}=\triangle_{2}$ and $a_{0}=0.11$. The blocklength
of polar codes is given by $N=2^{12},2^{14},2^{16},2^{18},2^{20}.$
\label{fig:Simulation_Lossy_DSBS_Region2}}
\end{figure}

Notice that the test channel is asymmetric, which is different from
the symmetric test channel $\left(\ref{eq:Transition probability}\right)$
for Curve GB. Therefore we should design polar codes for the lossy
compression with asymmetric test channels, as mentioned in Section
\ref{sub:PCforLossyCompression}. After that, we send the compressed
sequence through the shared branch to the two decoders. The simulation
performance for $\left(\triangle_{1},\triangle_{2}\right)\in\varepsilon_{2}$
is shown in Fig. \ref{fig:Simulation_Lossy_DSBS_Region2}. The dashed
line is the achievable bound $R_{XY}\left(\triangle_{1},\triangle_{2}\right)$
when $\left(\triangle_{1},\triangle_{2}\right)\in\varepsilon_{2}$
and $\triangle_{1}=\triangle_{2}$. As for the lines of simulation
results with $N=2^{12},2^{14},2^{16},2^{18},2^{20}$, the horizontal
axis refers to the average distortion between the practical $\triangle_{1}$
and $\triangle_{2}$. In fact, the practical $\triangle_{1}$ and
$\triangle_{2}$ tend to be very close to each other when the number
of simulation rounds is large.

Region $\varepsilon_{3}$ is a degenerate region in the sense that,
for example, if $a_{0}<\frac{\triangle_{2}-\triangle_{1}}{1-2\triangle_{1}}$
and $\triangle_{i}\leq\frac{1}{2}$, $i=1,2$, we have $R_{XY}\left(\triangle_{1},\triangle_{2}\right)=R_{X}\left(\triangle_{1}\right)$.
This implies that the optimal code strategy is to ignore $Y$ and
optimally compress $X$. Hence, $Y'$ can be estimated from $X'$
with distortion less than $\triangle_{2}$. The case when $a_{0}<\frac{\triangle_{1}-\triangle_{2}}{1-2\triangle_{2}}$
can be solved similarly. Therefore, polar codes can be designed for
the lossy compression of a single source in this region. 

So far, we have constructed polar codes to extract the lossy common
information $C\left(\triangle_{1},\triangle_{2}\right)$ in the entire
distortion region except for region $\varepsilon_{11}$. From \cite{GeXulossyCI},
we know that $C\left(X,Y\right)\leq C\left(\triangle_{1},\triangle_{2}\right)$
for $\left(\triangle_{1},\triangle_{2}\right)\in\varepsilon_{11}$;
however, the exact value of the lossy common information in region
$\varepsilon_{11}$ remains unknown.

\section{Lossy Gray-Wyner Coding for Gaussian RVs \label{sub:CI for joint Gaussian} }

Apart from the DSBS case addressed in the previous section, the lossy
common information has also been generalized to two Gaussian RVs in
\cite{GeXulossyCI}. In this section, we propose a coding scheme
to extract the lossy common information of a pair of joint Gaussian
sources using polar lattices. 

\begin{figure}
\begin{centering}
\includegraphics[scale=0.7]{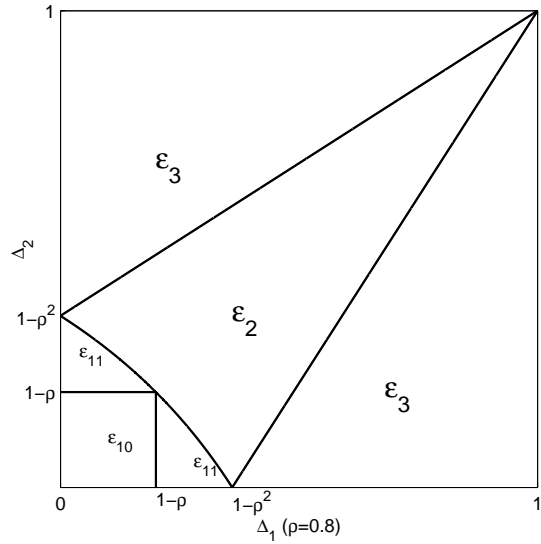}
\par\end{centering}

\protect\caption{The distortion regions for bivariate Gaussian RVs when $\rho=0.8$.
\label{fig:The-distortion-regions-GaussianRVs}}

\end{figure}

Let $X$, $Y$ be two Gaussian RVs with zero mean and covariance matrix
\[
K_{2}=\ensuremath{\begin{bmatrix}1 & \rho\\
\rho & 1
\end{bmatrix},}
\]
with $0<\rho<1$. The lossy common information for $\left(X,Y\right)$
has been given by \cite{GeXulossyCI} 
\begin{equation}
C\left(\triangle_{1},\triangle_{2}\right)=\begin{cases}
C\left(X,Y\right) & \left(\triangle_{1},\triangle_{2}\right)\in\varepsilon_{10},\\
R_{XY}\left(\triangle_{1},\triangle_{2}\right) & \left(\triangle_{1},\triangle_{2}\right)\in\varepsilon_{2}\cup\varepsilon_{3}\\
0 & \left(\triangle_{1},\triangle_{2}\right)\geq\left(1,1\right),
\end{cases},\label{eq:LossyCI_TwoGaussian}
\end{equation}

\[
C\left(X,Y\right)\leq C\left(\triangle_{1},\triangle_{2}\right)\leq R_{XY}\left(\triangle_{1},\triangle_{2}\right),\textrm{ }\textrm{ }\textrm{ }\left(\triangle_{1},\triangle_{2}\right)\in\varepsilon_{11},
\]
where 
\begin{align*}
\varepsilon_{10} & =\left\{ \left(\triangle_{1},\triangle_{2}\right):0\leq\triangle_{i}\leq1-\rho,\textrm{ }i=1,2\right\} ,\\
\varepsilon_{11} & =\varepsilon_{10}^{c}\cap\left\{ \left(\triangle_{1},\triangle_{2}\right):\triangle_{1}+\triangle_{2}-\triangle_{1}\triangle_{2}\leq1-\rho^{2}\right\} ,\\
\varepsilon_{2} & =\varepsilon_{10}^{c}\cap\varepsilon_{11}^{c}\cap\left\{ \left(\triangle_{1},\triangle_{2}\right):\min\left\{ \frac{1-\triangle_{1}}{1-\triangle_{2}},\frac{1-\triangle_{2}}{1-\triangle_{1}}\right\} \leq\rho^{2}\right\} ,\\
\varepsilon_{3} & =\varepsilon_{10}^{c}\cap\varepsilon_{11}^{c}\cap\varepsilon_{2}^{c}\cap\left\{ \left(\triangle_{1},\triangle_{2}\right):\triangle_{i}\leq1,\textrm{ }i=1,2\right\} .
\end{align*}
These distortion regions are illustrated in Fig. \ref{fig:The-distortion-regions-GaussianRVs}.
The joint rate-distortion function of the Gaussian sources $\left(X,Y\right)$
described above is given by \cite{nayak2010successive} 
\[
\begin{aligned} & R_{XY}\left(\triangle_{1},\triangle_{2}\right)=\begin{cases}
\frac{1}{2}\log\frac{1-\rho^{2}}{\triangle_{1}\triangle_{2}}, & \left(\triangle_{1},\triangle_{2}\right)\in\varepsilon_{1},\\
\frac{1}{2}\log\frac{1-\rho^{2}}{\triangle_{1}\triangle_{2}-\left(\rho-\sqrt{\left(1-\triangle_{1}\right)\left(1-\triangle_{2}\right)}\right)^{2}}, & \left(\triangle_{1},\triangle_{2}\right)\in\varepsilon_{2},\\
\frac{1}{2}\log\frac{1}{\min\left\{ \triangle_{1},\triangle_{2}\right\} }, & \left(\triangle_{1},\triangle_{2}\right)\in\varepsilon_{3},
\end{cases}\end{aligned}
\]
where $\varepsilon_{1}=\varepsilon_{10}\cup\varepsilon_{11}$. 

Notice that the relation $C\left(\triangle,\triangle\right)=C\left(X,Y\right)$
for $\triangle\leq1-\rho$ was firstly proposed in \cite{GeXulossyCI_conf}.
Then it has been extended to the case where $\triangle_{1}\neq\triangle_{2}$
as presented in (\ref{eq:LossyCI_TwoGaussian}).

Next we show how to extract the lossy common information that lies
in each of the distortion regions $\varepsilon_{10}$, $\varepsilon_{2}$
and $\varepsilon_{3}$. Notice that the characterizations of the common
RV $W$ defined in (\ref{eq:Character_lossyCI}) applies in $\varepsilon_{10}$,
$\varepsilon_{2}$ and $\varepsilon_{3}$. 

We propose a discretized version of $W$ to convey the lossy common
information of two joint Gaussian RVs, according to Wyner's second
approach to the characterization of common information \cite{WynerCI}.
The discretized version of $W$ is obtained by the use of polar lattices
\cite{PolarlatticeQZ,polarlatticeJ}. Some definitions that are necessary
for our scheme are given as follows.

An $n$-dimensional full-rank lattice is a discrete subgroup of $\mathbb{R}^{n}$
which can be defined by 
\[
\Lambda=\{\lambda=\mathbf{B}z:\textrm{ }z\in\mathbb{Z}^{n}\},
\]
where $\mathbf{B}$ is the $n\times n$ generator matrix. For $\sigma>0$
and $c\in\mathbb{R}^{n}$, the Gaussian distribution of variance $\sigma^{2}$
centered at $c$ is defined as 
\[
f_{\sigma,c}(x)=\frac{1}{(\sqrt{2\pi}\sigma)^{n}}e^{-\frac{\|x-c\|^{2}}{2\sigma^{2}}},\:\:x\in\mathbb{R}^{n}.
\]
Let $f_{\sigma,0}(x)=f_{\sigma}(x)$ for short. 

Define a $\Lambda$-periodic function 
\[
f_{\sigma,\Lambda}(x)=\sum\limits _{\lambda\in\Lambda}f_{\sigma,\lambda}(x)=\frac{1}{(\sqrt{2\pi}\sigma)^{n}}\sum\limits _{\lambda\in\Lambda}e^{-\frac{\|x-\lambda\|^{2}}{2\sigma^{2}}}.
\]
We note that $f_{\sigma,\Lambda}(x)$ is a probability density function
(PDF) if $x$ is restricted to the fundamental region $\mathcal{R}(\Lambda)$.
It is actually the PDF of the $\Lambda$-aliased Gaussian noise \cite{forney6}.

The flatness factor of a lattice $\Lambda$ is defined as \cite{forney6}
\begin{equation}
\epsilon_{\Lambda}(\sigma)\triangleq\max\limits _{x\in\mathcal{R}(\Lambda)}|V(\Lambda)f_{\sigma,\Lambda}(x)-1|,\label{eq:FlatnessFactor}
\end{equation}
where $V(\Lambda)=|\text{det}({B})|$ denotes the volume of a fundamental
region of $\Lambda$. It can be interpreted as the maximum variation
of $f_{\sigma,\Lambda}(x)$ with respect to the uniform distribution
over a fundamental region of $\Lambda$.

We define the discrete Gaussian distribution over $\Lambda$ centered
at $c$ as the discrete distribution taking values in $\lambda \in \Lambda$
as 
\[
D_{\Lambda,\sigma,c}(\lambda)=\frac{f_{\sigma,c}(\lambda)}{f_{\sigma,c}(\Lambda)},\;\forall\lambda\in\Lambda,
\]
 \noindent where $f_{\sigma,\mathrm{c}}(\Lambda)=\sum_{\lambda \in \Lambda} f_{\sigma,\mathrm{c}}(\lambda)$.
For convenience, we write $D_{\Lambda,\sigma}=D_{\Lambda,\sigma,\mathrm{0}}$.
It has been shown in \cite{LingBel13} that lattice Gaussian distribution
preserves many properties of the continuous Gaussian distribution
when the flatness factor is negligible. To keep the notations simple,
we always set $c=0$ and $n=1$ (one-dimensional lattice $\Lambda$)
in this work.

In addition, the Kullback-Leibler divergence of continuous distributions
$f_{X}$ and $f_{Y}$ is defined by

\[
\mathbb{D}(f_{X}\|f_{Y})=\int_{\mathbb{R}}f_{X}(x)\log\frac{f_{X}(x)}{f_{Y}(x)}dx.
\]
The variation distance is defined by 

\[
\mathbb{V}(f_{X}\|f_{Y})=\int_{\mathbb{R}}\left|f_{X}(x)-f_{Y}(x)\right|dx.
\]

\subsection{Lossy Common Information for Region $\varepsilon_{10}$}

For $\left(\triangle_{1},\triangle_{2}\right)\in\varepsilon_{10}$,
the lossy common information of $(X,Y)$ is conveyed by a Gaussian
RV $W$ with mean $0$ and variance $\rho$ such that 

\begin{equation}
\begin{cases}
X=W+\sqrt{1-\rho}N_{1}\\
Y=W+\sqrt{1-\rho}N_{2}
\end{cases}\label{eq:JointGaussianTestChannel}
\end{equation}
where $N_1$ and $N_2$ are standard Gaussian RVs and $N_{1},N_{2},W$
are independent of each other \cite{GeXulossyCI}. Clearly, the lossy
common information is given by 
\[
I\left(X,Y;W\right)=\frac{1}{2}\log\frac{1+\rho}{1-\rho}.
\]

\begin{lem}
\label{lem:Lemma Gaussian}Let $\bar{W}$ be a RV which follows a
discrete Gaussian distribution $D_{\Lambda,\sqrt{\rho}}$. Consider
two continuous RVs $\bar{X}$ and $\bar{Y}$

\[
\begin{cases}
\bar{X}=\bar{W}+\sqrt{1-\rho}N_{1}\\
\bar{Y}=\bar{W}+\sqrt{1-\rho}N_{2}
\end{cases}
\]
where $N_1$ and $N_2$ are the same as that given in (\ref{eq:JointGaussianTestChannel}).
Let $f_{\bar{X},\bar{Y}}(x,y)$ and $f_{X,Y}(x,y)$ denote the joint
PDF of $(\bar{X},\bar{Y})$ and $(X,Y)$, respectively. If $\epsilon=\epsilon_{\Lambda}\Big(\sqrt{\frac{\rho(1-\rho)}{1+\rho}}\Big)<\frac{1}{2}$,
the variation distance between $f_{\bar{X},\bar{Y}}(x,y)$ and $f_{X,Y}(x,y)$
is upper-bounded by 

\[
\int_{\mathbb{R}^{2}}\left|f_{\bar{X},\bar{Y}}(x,y)-f_{X,Y}(x,y)\right|dxdy\leq4\epsilon,
\]
and the mutual information $I(\bar{X},\bar{Y};\bar{W})$ satisfies

\[
I(\bar{{X}},\bar{{Y}};\bar{{W}})\geq\frac{1}{2}\log\frac{1+\rho}{1-\rho}-5\epsilon\log(e).
\]
According to Wyner's second approach, $\bar{W}$ is an eligible candidate
of the common message of $(X,Y)$ when $\epsilon \to 0$.\end{lem}
\begin{IEEEproof}
Since $\bar{X} - \bar{W} - \bar{Y}$ is a Markov chain, we have 

\begin{equation}
\begin{aligned} & f_{\bar{X},\bar{Y}}(x,y)\\
 & =\sum_{a\in\Lambda}f_{\bar{X},\bar{Y},\bar{W}}(x,y,a)\\
 & =\sum_{a\in\Lambda}f_{\bar{W}}(a)f_{\bar{X}|\bar{W}}(x|a)f_{\bar{Y}|\bar{W}}(y|a)\\
 & =\frac{1}{f_{\sqrt{\rho}}(\Lambda)}\sum_{a\in\Lambda}\frac{1}{\sqrt{2\pi\rho}}\exp\bigg(-\frac{a^{2}}{2\rho}\bigg)\frac{1}{\sqrt{2\pi(1-\rho)}}\exp\bigg(-\frac{(x-a)^{2}}{2(1-\rho)}\bigg)\frac{1}{\sqrt{2\pi(1-\rho)}}\exp\bigg(-\frac{(y-a)^{2}}{2(1-\rho)}\bigg)\\
 & =\frac{1}{2\pi\sqrt{1-\rho^{2}}}\exp\bigg(-\frac{x^{2}+y^{2}-2\rho xy}{2(1-\rho^{2})}\bigg)\frac{\sum_{a\in\Lambda}\frac{1}{\sqrt{2\pi\frac{\rho(1-\rho)}{1+\rho}}}\exp\bigg(-\frac{(a-\frac{\rho(x+y)}{1+\rho})^{2}}{2\cdot\frac{\rho(1-\rho)}{1+\rho}}\bigg)}{f_{\sqrt{\rho}}(\Lambda)},
\end{aligned}
\label{eq:GaussianJointProbXY}
\end{equation}
 where $\frac{1}{2\pi\sqrt{1-\rho^2}} \exp \bigg( -\frac{x^2+y^2-2\rho xy}{2(1-\rho^2)} \bigg)=f_{X,Y}(x,y)$
is the PDF of two joint Gaussian RVs. By the definition of the flatness
factor (\ref{eq:FlatnessFactor}), we have 

\begin{equation}
\begin{aligned}\left|V(\Lambda)\sum_{a\in\Lambda}\frac{1}{\sqrt{2\pi\frac{\rho(1-\rho)}{1+\rho}}}\exp\bigg(-\frac{(a-\frac{\rho(x+y)}{1+\rho})^{2}}{2\frac{\rho(1-\rho)}{1+\rho}}\bigg)-1\right|\leq\epsilon_{\Lambda}\Bigg(\sqrt{\frac{\rho(1-\rho)}{1+\rho}}\Bigg)=\epsilon.\end{aligned}
\label{eq:GaussianFlatFactor}
\end{equation}
Since $\epsilon_{\Lambda}(\sigma)$ is a monotonically decreasing
function of $\sigma$ (see \cite[Remark 2]{cong2}), we have $\epsilon(\sqrt{\rho}) \leq \epsilon$
and hence 

\begin{equation}
\left|V(\Lambda)f_{\sqrt{\rho}}(\Lambda)-1\right|\leq\epsilon.\label{eq:GaussianFlatFactor2}
\end{equation}

Combining (\ref{eq:GaussianJointProbXY}), (\ref{eq:GaussianFlatFactor})
and (\ref{eq:GaussianFlatFactor2}) gives us

\[
f_{X,Y}(x,y)(1-2\epsilon)\leq f_{X,Y}(x,y)\cdot\frac{1-\epsilon}{1+\epsilon}\leq f_{\bar{X},\bar{Y}}(x,y),
\]
and

\[
f_{\bar{X},\bar{Y}}(x,y)\leq f_{X,Y}(x,y)\cdot\frac{1+\epsilon}{1-\epsilon}\leq f_{X,Y}(x,y)(1+4\epsilon),
\]
when $\epsilon<\frac{1}{2}$. Finally,

\begin{align*}
\int_{\mathbb{R}^{2}}\left|f_{\bar{X},\bar{Y}}(x,y)-f_{X,Y}(x,y)\right|dxdy\leq4\epsilon\int_{\mathbb{R}^{2}}f_{X,Y}(x,y)dxdy=4\epsilon.
\end{align*}

Similarly, the Kullback-Leibler divergence between $f_{\bar{X},\bar{Y}}(x,y)$
and $f_{X,Y}(x,y)$ can be upper-bounded as 

\begin{equation}
\begin{aligned}\mathbb{D}(f_{\bar{X},\bar{Y}}\|f_{X,Y}) & =\int_{\mathbb{R}^{2}}f_{\bar{X},\bar{Y}}(x,y)\log\frac{f_{\bar{X},\bar{Y}}(x,y)}{f_{X,Y}(x,y)}dxdy\\
 & \leq\int_{\mathbb{R}^{2}}f_{\bar{X},\bar{Y}}(x,y)\log(1+4\epsilon)dxdy\\
 & =\log(1+4\epsilon).
\end{aligned}
\label{eq:GaussianDistance}
\end{equation}

\noindent For any $\sqrt{\frac{\rho(1-\rho)}{1+\rho}}>0$, $\epsilon$
can be made arbitrarily small by scaling $\Lambda$. Therefore, when
$\epsilon \to 0$, $\bar{W}$ can be viewed as the common message
according to Wyner's second approach. To see that $I(\bar{X},\bar{Y};\bar{W})$
can be arbitrarily close to the lossy common information, we rewrite
$\mathbb{D}(f_{\bar{X},\bar{Y}}\|f_{X,Y})$ as

\[
\begin{aligned} & \mathbb{D}(f_{\bar{X},\bar{Y}}\|f_{X,Y})\\
 & =\int_{\mathbb{R}^{2}}f_{\bar{X},\bar{Y}}(x,y)\log\frac{f_{\bar{X},\bar{Y}}(x,y)}{f_{X,Y}(x,y)}dxdy\\
 & =-\int_{\mathbb{R}^{2}}f_{\bar{X},\bar{Y}}(x,y)\log f_{X,Y}(x,y)dxdy-h(\bar{X},\bar{Y})\\
 & =-\int_{\mathbb{R}^{2}}f_{\bar{X},\bar{Y}}(x,y)\log\Bigg(\frac{1}{2\pi\sqrt{1-\rho^{2}}}\exp\bigg(-\frac{x^{2}+y^{2}-2\rho xy}{2(1-\rho^{2})}\bigg)\Bigg)dxdy-h(\bar{X},\bar{Y})\\
 & =\log\big(2\pi\sqrt{1-\rho^{2}}\big)+\frac{\mathsf{E}_{\bar{X},\bar{Y}}[x^{2}+y^{2}-2\rho xy]}{2(1-\rho^{2})}\log(e)-h(\bar{X},\bar{Y})\\
 & =\log\big(2\pi\sqrt{1-\rho^{2}}\big)+\frac{1+\mathsf{E}_{\bar{W}}[w^{2}]}{1+\rho}\log(e)-h(\bar{X},\bar{Y}).
\end{aligned}
\]

\noindent Note that $\epsilon_{\Lambda}(\sqrt{\rho})\leq \epsilon$.
By \cite[Lemma 5]{LingBel13} and \cite[Remark 3]{LingBel13} ,
it is easy to make $\mathsf{E}_{\bar{W}}\left[w^{2}\right]\geq\rho(1-2\epsilon)$.
Then we have

\[
\begin{aligned}\mathbb{D}(f_{\bar{X},\bar{Y}}\|f_{X,Y}) & \geq\log\big(2\pi\sqrt{1-\rho^{2}}\big)+(1-\epsilon)\log(e)-h(\bar{X},\bar{Y})\\
 & =h(X,Y)-h(\bar{X},\bar{Y})-\epsilon\log(e).
\end{aligned}
\]

Using (\ref{eq:GaussianDistance}), we obtain

\[
\begin{aligned}I(X,Y;W)-I(\bar{X},\bar{Y};\bar{W}) & =h(X,Y)-h(\bar{X},\bar{Y})\\
 & \leq\log(1+4\epsilon)+\epsilon\log(e)\\
 & \leq5\epsilon\log(e).
\end{aligned}
\]

\end{IEEEproof}
Similar to \cite{PolarlatticeQZ}, using $D_{\Lambda,\sqrt{\rho}}$
as the reconstruction distribution, we can design a quantization polar
lattice from ``Construction D'' to extract the lossy common information.
The only difference is that the size of the source alphabet is doubled
in this work. The next theorem shows that the design of polar lattices
for extracting the lossy common information of a pair of joint Gaussian
sources is exactly the same as that for quantizing a single Gaussian
source, which means that the technique proposed in \cite{PolarlatticeQZ}
can be directly employed to our work.
\begin{thm}
\label{thm:GaussianSource}The construction of a polar lattice for
extracting the lossy common information of a pair of joint Gaussian
sources $(X,Y)$ in distortion region $\varepsilon_{10}$ is equivalent
to the construction of a rate-distortion bound achieving polar lattice
for a Gaussian source $\frac{X+Y}{2}$.\end{thm}
\begin{IEEEproof}
Let $\bar{W}$ be labeled by bits $\bar{W}_{1}^{r}=\{\bar{W}_{1},\cdots,\bar{W}_{r}\}$
according to a binary partition chain $\Lambda/\Lambda_{1}/\cdots/\Lambda_{r-1}/\Lambda'$
($\Lambda'$ also refers to $\Lambda_{r}$). Then, $D_{\Lambda,\sqrt{\rho}}$
induces a distribution $P_{\bar{W}_{1}^{r}}$ whose limit corresponds
to $D_{\Lambda,\sqrt{\rho}}$ as ${r\rightarrow\infty}$. 

By the chain rule of mutual information,

\[
I(\bar{X},\bar{Y};\bar{W}_{1}^{r})=\sum_{\ell=1}^{r}I(\bar{X},\bar{Y};\bar{W}_{\ell}|\bar{W}_{1}^{\ell-1}),
\]

\noindent we obtain $r$ binary-input test channels ${V}_{\ell}$
for $1\leq \ell \leq r$. Given the realization $w_{1}^{\ell-1}$
of $\bar{W}_{1}^{\ell-1}$, denote by $\mathcal{A}_{\ell}(w_{1}^{\ell})$
the coset of $\Lambda_{\ell}$ indexed by $w_{1}^{\ell-1}$ and $w_{\ell}$.
According to \cite{multilevel}, the channel transition PDF of the
$\ell$-th channel ${V}_{\ell}$ is given by 

\[
\begin{aligned} & f_{\bar{X},\bar{Y}|\bar{W}_{\ell},\bar{W}_{1}^{\ell-1}}(x,y|w_{\ell},w_{1}^{\ell-1})\\
 & =\frac{1}{f_{\sqrt{\rho}}(\mathcal{A}_{\ell}(w_{1}^{\ell}))}\sum_{a\in\mathcal{A}_{\ell}(w_{1}^{\ell})}f_{\sqrt{\rho}}(a)f_{\bar{X},\bar{Y}|\bar{W}}(x,y|a)\\
 & =\frac{1}{f_{\sqrt{\rho}}(\mathcal{A}_{\ell}(w_{1}^{\ell}))}\sum_{a\in\mathcal{A}_{\ell}(w_{1}^{\ell})}\frac{1}{\sqrt{2\pi\rho}}\exp\bigg(-\frac{a^{2}}{2\rho}\bigg)\frac{1}{\sqrt{2\pi(1-\rho)}}\exp\bigg(-\frac{(x-a)^{2}}{2(1-\rho)}\bigg)\frac{1}{\sqrt{2\pi(1-\rho)}}\exp\bigg(-\frac{(y-a)^{2}}{2(1-\rho)}\bigg)\\
 & =\frac{1}{2\pi\sqrt{1-\rho^{2}}}\exp\bigg(-\frac{x^{2}+y^{2}-2\rho xy}{2(1-\rho^{2})}\bigg)\frac{1}{f_{\sqrt{\rho}}(\mathcal{A}_{\ell}(w_{1}^{\ell}))}\sum_{a\in\mathcal{A}_{\ell}(w_{1}^{\ell})}\frac{1}{\sqrt{2\pi\frac{\rho(1-\rho)}{1+\rho}}}\exp\bigg(-\frac{(a-\frac{\rho(x+y)}{1+\rho})^{2}}{2\cdot\frac{\rho(1-\rho)}{1+\rho}}\bigg).
\end{aligned}
\]

Let $\tilde{V}_{l}$ be a symmetrized channel with input $\tilde{W}_{l}$
(assume to be uniformly distributed) and output $\left(\bar{X},\bar{Y},\bar{W}_{1}^{\ell-1},\bar{W}_{l}\oplus\tilde{W}_{l}\right)$,
built from the asymmetric channel $V_{l}$. Then the joint PDF of
$V_{l}$ can be represented by the transition PDF of $\tilde{V}_{l}$
(see \cite{polarlatticeJ} for more details), as shown in the following
equation. 

\begin{equation}
\begin{aligned} & f_{\tilde{V}_{\ell}}(x,y,w_{1}^{\ell-1},w_{\ell}\oplus\tilde{w}_{\ell}|\tilde{w}_{\ell})\\
 & =f_{\bar{X},\bar{Y},\bar{W}_{1}^{\ell}}(x,y,w_{1}^{\ell})\\
 & =\frac{1}{2\pi\sqrt{1-\rho^{2}}}\exp\bigg(-\frac{x^{2}+y^{2}-2\rho xy}{2(1-\rho^{2})}\bigg)\frac{1}{f_{\sqrt{\rho}}(\Lambda)}\sum_{a\in\mathcal{A}_{\ell}(w_{1}^{\ell})}\frac{1}{\sqrt{2\pi\frac{\rho(1-\rho)}{1+\rho}}}\exp\bigg(-\frac{(a-\frac{\rho(x+y)}{1+\rho})^{2}}{2\cdot\frac{\rho(1-\rho)}{1+\rho}}\bigg).
\end{aligned}
\label{eq:GaussianSymmetrimizedF}
\end{equation}

Comparing with the $\Lambda_{\ell-1}/\Lambda_{\ell}$ channel \cite[ Equation (13)]{polarlatticeJ},
it can be derived that the symmetrized channel (\ref{eq:GaussianSymmetrimizedF})
is equivalent to a $\Lambda_{\ell-1}/\Lambda_{\ell}$ channel with
noise variance $\frac{\rho(1-\rho)}{1+\rho}$. To construct polar
lattices, we are interested in the likelihood ratio derived by (\ref{eq:GaussianSymmetrimizedF}).
Moreover, the likelihood ratios are affected by the summation section
at the end of (\ref{eq:GaussianSymmetrimizedF}). Fortunately, we
have found an easier way to achieve the same likelihood ratios by
quantizing a single Gaussian source $\frac{X+Y}{2}$ using the reconstruction
distribution $D_{\Lambda,\sqrt{\rho}}$. Follows are the explanations.

Recall that $X$, $Y$ are bivariate Gaussian with zero mean and covariance
matrix $\ensuremath{\left[\begin{smallmatrix}1 & \rho\\
\rho & 1
\end{smallmatrix}\right]}$. Therefore, $\frac{X+Y}{2}$ is Gaussian with zero mean and variance
\[
\begin{aligned}\sigma^{2}\left(\frac{X+Y}{2}\right) & \mathsf{=E}\left[\left(\frac{X+Y}{2}\right)^{2}\right]-\left(\mathsf{E}\left[\frac{X+Y}{2}\right]\right)^{2}\\
 & =\frac{1}{4}\left(\mathsf{E}\left[X^{2}+Y^{2}+2XY\right]\right)-0\\
 & =\frac{1+\rho}{2}.
\end{aligned}
\]
Let us consider the construction of a polar lattice to quantize $\frac{X+Y}{2}$
using the reconstruction distribution $D_{\Lambda, \sqrt{\rho}}$.
Denote the variance of the source and the reconstruction as $\sigma_{s}^{2}=\frac{1+\rho}{2}$
and $\sigma_{r}^{2}=\rho$, respectively. Thus, the variance of the
noise equals $\sigma_{z}^{2}=\sigma_{s}^{2}-\sigma_{r}^{2}=\frac{1-\rho}{2}$.
Then we perform Minimum Mean Square Error (MMSE) rescaling on this
relation. By definitions, the MMSE scaling coefficient $\alpha$ and
noise variance $\tilde{\sigma}_{z}^{2}$ are given by 
\[
\alpha=\frac{\sigma_{r}^{2}}{\sigma_{s}^{2}}=\frac{2\rho}{1+\rho},
\]

\[
\tilde{\sigma}_{z}^{2}=\alpha\cdot\sigma_{z}^{2}=\frac{\rho(1-\rho)}{1+\rho},
\]
which are the same as those in the summation section of (\ref{eq:GaussianSymmetrimizedF}). 
\end{IEEEproof}
The result of Lemma \ref{lem:Lemma Gaussian} and Theorem \ref{thm:GaussianSource}
can be generalized to multivariate Gaussian sources presented in Section
\ref{sub:MultiGaussian}. The common message of multivariate Gaussian
RVs can also be conveyed by a discretized RV. Moreover, the construction
of polar lattices can be designed in the same way as that of a single
Gaussian source, given by the arithmetic mean of multiple Gaussian
sources.

So far, we have presented how to extract the lossy common information
for region $\varepsilon_{10}$. Next we show how to achieve the distortions
$\left(\triangle_{1},\triangle_{2}\right)\in\varepsilon_{10}$ from
the extracted $\bar{W}$ and the Gaussian sources $\left(X,Y\right)$. 

Firstly, the conditional rate-distortion function $R_{X|W}\left(\triangle_{1}\right)$
is defined by \cite{GrayConditionalRD} 
\[
R_{X|W}\left(\triangle_{1}\right)=\min_{P_{\left(X'|X,W\right)}\left(x'|x,w\right):\mathsf{E}d(X',X)\leq\triangle_{1}}I\left(X;X'|W\right).
\]
In region $\varepsilon_{10}$, the conditional distribution of $X$
given $W$ is a Gaussian distribution with variance $1-\rho$ from
the test channel $\left(\ref{eq:JointGaussianTestChannel}\right)$,
therefore 
\[
\begin{cases}
R_{X|W}\left(\triangle_{1}\right)=\frac{1}{2}\log\frac{1-\rho}{\triangle_{1}}\\
R_{Y|W}\left(\triangle_{2}\right)=\frac{1}{2}\log\frac{1-\rho}{\triangle_{2}}.
\end{cases}
\]
Hence the condition 
\[
R_{X|W}\left(\triangle_{1}\right)+R_{Y|W}\left(\triangle_{2}\right)+I\left(X,Y;W\right)=R_{XY}\left(\triangle_{1},\triangle_{2}\right)
\]
is satisfied \cite{GeXulossyCI} for region $\varepsilon_{10}$. 

Notice that $X-\bar{W}$ and $\bar{X}-\bar{W}$ can be made arbitrarily
close to each other, since $\mathbb{V}\left(f_{X}\left(x\right),f_{\bar{X}}\left(x\right)\right)\leq4\epsilon$
\cite{LingBel13} and $\epsilon$ can be scaled very close to zero.
The difference between $\bar{X}$ and $\bar{W}$ can be regarded another
Gaussian source $\bar{X}-\bar{W}=\sqrt{1-\rho}N_{1}$ from $\left(\ref{eq:JointGaussianTestChannel}\right)$.
Then we apply the lossy compression using polar lattices \cite{PolarlatticeQZ}
to the source $\sqrt{1-\rho}N_{1}$ with distortion $\triangle_{1}$.
As a result, the reconstruction RV can be represented as $\sqrt{1-\rho-\triangle_{1}}\bar{N}$,
where $\bar{N}$ follows a discrete Gaussian distribution. Next we
shall use $\sqrt{1-\rho-\triangle_{1}}\bar{N}$ to reconstruct $\bar{X'}$
through either the shared channel or the private channel. More explicitly,
at the decoder, the reconstructed $\bar{X'}$ can be derived by 
\[
\bar{X'}=\bar{W}+\sqrt{1-\rho-\triangle_{1}}\bar{N}.
\]
The distortion between $X$ and $\bar{X'}$ is approaching $\triangle_{1}$,
when the compression rate $R>\frac{1}{2}\log\frac{1-\rho}{\triangle_{1}}$,
$N\rightarrow\infty$ and $\epsilon\rightarrow0$. Similarly, the
distortion $\triangle_{2}$ of the source $Y$ can be derived.

\subsection{Common Information for Multiple Joint Gaussian Sources\label{sub:MultiGaussian}}

First, we define the common information of $L$ dependent RVs. Let
$\mathbf{X}_{L}\triangleq\{X_{1},X_{2},\ldots X_{L}\}$ be $L$ dependent
RVs that take values in some arbitrary space $\mathcal{X}_{1}\times\mathcal{X}_{2}\times\cdots\times\mathcal{X}_{L}$.
The joint distribution of $\mathbf{X}_{L}$ is denoted by $P_{\mathbf{X}_{L}}(\mathbf{x}_{L})$,
which is either a probability mass function or a PDF.
\begin{defn}
The Wyner's common information of multiple Gaussian sources $\mathbf{X}_{L}$
has been defined in \cite{GeXulossyCI} as follows, 
\[
C\left(\mathbf{X}_{L}\right)\triangleq\inf I\left(\mathbf{X}_{L};W\right),
\]

\end{defn}
\noindent where the infimum is taken over all the joint distributions
of $\left(\mathbf{X}_{L},W\right)$ such that 
\begin{itemize}
\item the marginal distribution for $\mathbf{X}_{L}$ is $P_{\mathbf{X}_{L}}(\mathbf{x}_{L})$,
\item $\mathbf{X}_{L}$ are conditionally independent given $W$. 
\end{itemize}
Then we show the construction of polar codes to extract the Wyner's
common information of multiple joint Gaussian sources.

For $L$ joint Gaussian RVs $\mathbf{X}_{L}=\{X_{1},X_{2},\ldots X_{L}\}$
with covariance matrix,

\begin{equation}
K_{L}=\begin{bmatrix}1 & \rho & \cdots & \rho\\
\rho & 1 & \cdots & \rho\\
\vdots & \vdots & \ddots & \vdots\\
\rho & \rho & \cdots & 1
\end{bmatrix}.\label{eq:covariance matrix KL}
\end{equation}

The common RV of $\mathbf{X}_{L}$ is conveyed by a Gaussian RV $W$
with mean $0$ and variance $\rho$ such that 
\begin{equation}
X_{i}=W+\sqrt{1-\rho}N_{i},\textrm{ }\label{eq:N gaussian TC}
\end{equation}

\noindent where $i=1,2,\ldots,L$. Besides, $N_{i}$ are standard
Gaussian RVs independent of each other and $W$. The common information
is given by \cite{GeXulossyCI} as 
\[
I(\mathbf{X}_{L};W)=\frac{1}{2}\log\left(1+\frac{L\rho}{1-\rho}\right).
\]

Similar to the problem where there are two joint Gaussian sources,
we apply a polar lattice to derive a discrete version of $W$ to represent
the common message of multiple Gaussian sources. The next lemma indicates
that the common information of $\bar{W}$ with discrete Gaussian distribution
is very close to the common information of $W$ with continuous Gaussian
distribution, when the flatness factor is negligible.
\begin{lem}
\label{lem:Lemma L Gaussian source }Let $\bar{W}$ be a RV which
follows a discrete Gaussian distribution $D_{\Lambda,\sqrt{\rho}}$.
Consider $L$ continuous RVs $\mathbf{\bar{X}}_{L}=\{X_{1},X_{2},\ldots X_{L}\}$
and relations 
\[
\bar{X}_{i}=\bar{W}+\sqrt{1-\rho}N_{i},\textrm{ }i=1,2,\ldots L,
\]
where $N_{i}$ are the same as that given in (\ref{eq:N gaussian TC}).
Let $f_{\mathbf{\bar{X}}_{L}}(\mathbf{x}_{L})$ and $f_{\mathbf{X}_{L}}(\mathbf{x}_{L})$
denote the joint PDF of $\mathbf{\bar{X}}_{L}=\{\bar{X}_{1},\bar{X}_{2},\ldots\bar{X}_{L}\}$
and $\mathbf{X}_{L}=\{X_{1},X_{2},\ldots X_{L}\}$, respectively.
If $\epsilon=\epsilon_{\Lambda}\left(\sqrt{\frac{\rho(1-\rho)}{1+(L-1)\rho}}\right)<\frac{1}{2}$,
the variation distance between $f_{\mathbf{\bar{X}}_{L}}(\mathbf{x}_{L})$
and $f_{\mathbf{X}_{L}}(\mathbf{x}_{L})$ is upper-bounded by 
\[
\int_{\mathbb{R}^{L}}\left|f_{\mathbf{\bar{X}}_{L}}(\mathbf{x}_{L})-f_{\mathbf{X}_{L}}(\mathbf{x}_{L})\right|dx_{1}dx_{2}\ldots dx_{L}\leq4\epsilon,
\]

\noindent and the mutual information $I(\mathbf{\bar{X}}_{L};\bar{W})$
satisfies 
\[
I(\mathbf{\bar{X}}_{L};\bar{W})\geq\frac{1}{2}\log\left(1+\frac{L\rho}{1-\rho}\right)-5\epsilon\log\left(e\right).
\]

\noindent according to Wyner's second approach, $\bar{W}$ is an
eligible candidate of the common message of \textup{$\mathbf{X}_{L}=\{X_{1},X_{2},\ldots X_{L}\}$
when $\epsilon\rightarrow0$.}\end{lem}
\begin{IEEEproof}
See Appendix \ref{sec:Proof-of-Lemma-LGaussian}.
\end{IEEEproof}
Next we use $D_{\Lambda,\sqrt{\rho}}$ as the reconstruction distribution
to design a polar lattice to extract the common information among
$L$ joint Gaussian sources. However, the construction will become
complicated when the number of sources is large. This problem can
be resolved by a similar scheme to the previous case, where the design
for two Gaussian sources can be reduced to that for a single Gaussian
source. The next theorem indicates that the construction to extract
Wyner's common information of multiple joint Gaussian sources is the
same as that for a single Gaussian source. Similarly, the technique
of quantization using polar lattices in \cite{PolarlatticeQZ} can
be directly employed to this case.
\begin{thm}
\label{thm:Theorem L Gaussian source}The construction of a polar
lattice for extracting the common information of $L$ joint Gaussian
sources $\mathbf{X}_{L}=\{X_{1},X_{2},\ldots X_{L}\}$ is equivalent
to the construction of a rate-distortion bound achieving polar lattice
for a Gaussian source $\frac{X_{1}+X_{2}+\ldots+X_{L}}{L}$.\end{thm}
\begin{IEEEproof}
See Appendix \ref{sec:Proof-of-Theorem-LGaussian}.
\end{IEEEproof}

\subsection{Lossy Common Information for Region $\varepsilon_{2}\cup\varepsilon_{3}$}

For region $\varepsilon_{2}$, the lossy common information of $\left(X,Y\right)$
equals the optimal rate for a certain distortion pair of the joint
Gaussian sources. It has been shown in \cite{GeXulossyCI} that the
$W$ satisfying (\ref{eq:Character_lossyCI}) supports the result
that 
\[
I\left(X,Y;W\right)=I\left(X,Y;X'\right)=I\left(X,Y;X',Y'\right),
\]
where $\left(X',Y'\right)$ achieve $R_{XY}\left(\triangle_{1},\triangle_{2}\right)$.
Therefore, the extraction of the lossy common information can be regarded
the lossy compression that achieves the joint rate-distortion bound
$R_{XY}\left(\triangle_{1},\triangle_{2}\right)$ of two correlated
Gaussian sources with zero-mean and covariance matrix $K_{2}$. Authors
in \cite{nayak2010successive} proposed an optimal backward test channel
for region $\varepsilon_{2}$, which is given by 
\begin{equation}
\begin{bmatrix}X\\
Y
\end{bmatrix}=\begin{bmatrix}X'\\
Y'
\end{bmatrix}+\begin{bmatrix}Z_{1}\\
Z_{2}
\end{bmatrix},\label{eq:TwoGaussianRegion2}
\end{equation}
where both $\left[X',Y'\right]^{\mathrm{T}}$ and $\left[Z_{1},Z_{2}\right]^{\mathrm{T}}$
are Gaussian vectors independent of each other and their covariance
matrices are respectively given by 
\[
\begin{aligned}K_{X',Y'} & =\begin{bmatrix}\delta_{1} & \sqrt{\delta_{1}\delta_{2}}\\
\sqrt{\delta_{1}\delta_{2}} & \delta_{2}
\end{bmatrix},\\
K_{Z_{1},Z_{2}} & =\begin{bmatrix}\triangle_{1} & \rho-\sqrt{\delta_{1}\delta_{2}}\\
\rho-\sqrt{\delta_{1}\delta_{2}} & \triangle_{2}
\end{bmatrix},
\end{aligned}
\]
for $\left(\triangle_{1},\triangle_{2}\right)\in\varepsilon_{2}$.
We use the notation 
\[
\delta_{i}\triangleq1-\triangle_{i},\textrm{ }i=1,2.
\]
Since $K_{X',Y'}$ is singular in this region, the relation between
$X'$ and $Y'$ is 
\[
Y'=\sqrt{\frac{\delta_{2}}{\delta_{1}}}X'.
\]

Let $\bar{X'}$ follow a discrete Gaussian distribution $D_{\Lambda,\sqrt{\delta_{1}}}$
, and $\bar{Y'}$ follow a discrete Gaussian distribution $D_{\Lambda,\sqrt{\delta_{2}}}$.
The covariance matrix of $\left(\bar{X'},\bar{Y'}\right)$ is the
same as $K_{X',Y'}$. Therefore $\left(\bar{X'},\bar{Y'}\right)$
also has the relation $\bar{Y'}=\sqrt{\frac{\delta_{2}}{\delta_{1}}}\bar{X'}$.
\begin{lem}
\label{lem:TwoGaussianRegion2}Consider two continuous RVs $\bar{X}$
and $\bar{Y}$ 
\[
\begin{bmatrix}\bar{X}\\
\bar{Y}
\end{bmatrix}=\begin{bmatrix}\bar{X'}\\
\bar{Y'}
\end{bmatrix}+\begin{bmatrix}Z_{1}\\
Z_{2}
\end{bmatrix},
\]
where $Z_{1}$ and $Z_{2}$ are the same as that given in (\ref{eq:TwoGaussianRegion2}).
Let $f_{\bar{X},\bar{Y}}(x,y)$ and $f_{X,Y}(x,y)$ denote the joint
PDF of $(\bar{X},\bar{Y})$ and $(X,Y)$, respectively. If $\epsilon=\epsilon_{\Lambda}\left(\sqrt{\frac{\delta_{1}\left(\triangle_{1}\triangle_{2}-\left(\rho-\sqrt{\delta_{1}\delta_{2}}\right)^{2}\right)}{1-\rho^{2}}}\right)<\frac{1}{2}$,
the variation distance between $f_{\bar{X},\bar{Y}}(x,y)$ and $f_{X,Y}(x,y)$
is upper-bounded by 

\[
\int_{\mathbb{R}^{2}}\left|f_{\bar{X},\bar{Y}}(x,y)-f_{X,Y}(x,y)\right|dxdy\leq4\epsilon,
\]
and the mutual information $I(\bar{X},\bar{Y};\bar{X'})$ satisfies

\[
I(\bar{{X}},\bar{{Y}};\bar{X'})\geq I(X,Y;X')-5\epsilon\log(e).
\]
$\bar{X'}$ is an eligible candidate of the common message of $(X,Y)$
when $\epsilon \to 0$.\end{lem}
\begin{IEEEproof}
See Appendix \ref{sec:Proof-of-Lemma-TwoGaussianRegion2}.\end{IEEEproof}
\begin{thm}
\label{thm:Gaussian_Region2_Them} Given two correlated Gaussian sources
$\left(X,Y\right)$ with zero mean and covariance matrix $K_{2}$
and an average distortion pair $\left(\triangle_{1},\triangle_{2}\right)\in\varepsilon_{2}$,
for any rate $R>\frac{1}{2}\log\frac{1-\rho^{2}}{\triangle_{1}\triangle_{2}-\left(\rho-\sqrt{\left(1-\triangle_{1}\right)\left(1-\triangle_{2}\right)}\right)^{2}}$,
there exists a polar lattice with rate $R$ such that the distortions
are arbitrarily close to $\left(\triangle_{1},\triangle_{2}\right)$
when $N\rightarrow\infty$, $r=\mathcal{O}\left(\log\log N\right)$
and the partition chain is scaled to make $\epsilon_{\Lambda}\left(\sqrt{\frac{\delta_{1}\left(\triangle_{1}\triangle_{2}-\left(\rho-\sqrt{\delta_{1}\delta_{2}}\right)^{2}\right)}{1-\rho^{2}}}\right)\rightarrow0$. \end{thm}
\begin{IEEEproof}
See Appendix \ref{sec:Proof-of-Theorem-Achivability}. \end{IEEEproof}
\begin{rem}
\label{rem:Region2}Similar to Theorem \ref{thm:GaussianSource},
the construction of a polar lattice for extracting the lossy common
information of a pair of joint Gaussian sources $\left(X,Y\right)$
in distortion region $\varepsilon_{2}$ is equivalent to the construction
of a rate-distortion bound achieving polar lattice for a Gaussian
source 
\[
U=\frac{\left(\delta_{1}-\rho\sqrt{\delta_{1}\delta_{2}}\right)X+\left(\sqrt{\delta_{1}\delta_{2}}-\rho\delta_{1}\right)Y}{\delta_{1}+\delta_{2}-2\rho\sqrt{\delta_{1}\delta_{2}}}.
\]

\end{rem}
This means that the lattice construction for a pair of Gaussian sources
is equivalent to that for a single source, which is simply a linear
combination of the two sources. 

It can be trivially derived that $U$ follows the Gaussian distribution
with zero mean and variance 
\[
\sigma^{2}\left[U\right]=\frac{\delta_{1}\left(1-\rho^{2}\right)}{\delta_{1}+\delta_{2}-2\rho\sqrt{\delta_{1}\delta_{2}}}.
\]

Consider the construction of a polar lattice to quantize $U$ using
the reconstruction distribution $D_{\Lambda,\sqrt{\delta_{1}}}$.
The MMSE coefficient and noise variance are respectively given by
\[
\begin{aligned}\alpha & =\frac{\delta_{1}+\delta_{2}-2\rho\sqrt{\delta_{1}\delta_{2}}}{1-\rho^{2}}\\
\sigma_{MMSE}^{2} & =\frac{\delta_{1}\left(\triangle_{1}\triangle_{1}-\left(\rho-\sqrt{\delta_{1}\delta_{2}}\right)^{2}\right)}{1-\rho^{2}}\\
 & =\delta_{1}-\frac{\delta_{1}\left(\delta_{1}+\delta_{2}-2\rho\sqrt{\delta_{1}\delta_{2}}\right)}{1-\rho^{2}}\\
 & =\delta_{1}-\sigma^{2}\left[\alpha U\right].
\end{aligned}
\]

Since the proof of Remark \ref{rem:Region2} follows quite similar
logic to that of Theorem \ref{thm:GaussianSource} based on Lemma
\ref{lem:TwoGaussianRegion2}, we will not include the proof in this
paper. 

\begin{figure}
\begin{centering}
\includegraphics[scale=0.9]{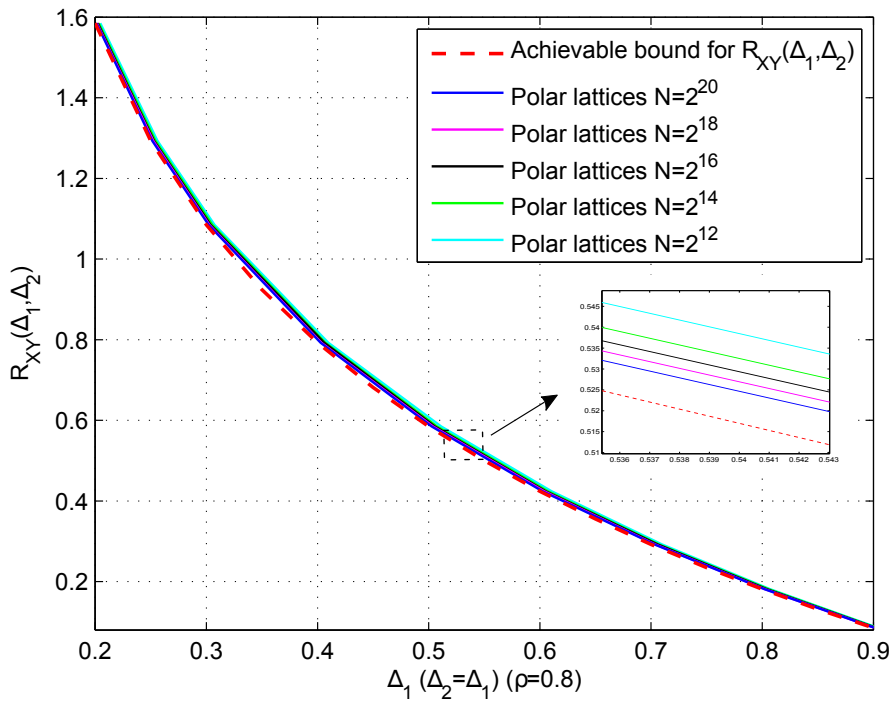}
\par\end{centering}

\protect\caption{Simulation performance for Gray-Wyner coding of two Gaussian sources
in region $\varepsilon_{2}$. We set $\triangle_{1}=\triangle_{2}$,
$\rho=0.8$. The blocklength of polar lattices of each level is given
by $N=2^{12}$, $2^{14}$, $2^{16}$, $2^{18}$, $2^{20}$. \label{fig:Simulation-performance-for-GaussianRVs}}

\end{figure}

The simulation results of region $\varepsilon_{2}$ are depicted in
Fig. \ref{fig:Simulation-performance-for-GaussianRVs}. The dashed
line is the achievable bound $R_{XY}\left(\triangle_{1},\triangle_{2}\right)$
when $\left(\triangle_{1},\triangle_{2}\right)\in\varepsilon_{2}$
and $\triangle_{1}=\triangle_{2}$. The correlation of Gaussian sources
$\left(X,Y\right)$ is set to $\rho=0.8$. Therefore, we have a wider
distortion range where $\triangle_{1}=\triangle_{2}\in\left(0.2,1\right)$.
As for the lines of simulation results with $N=2^{12},2^{14},2^{16},2^{18},2^{20}$,
the horizontal axis refers to the average distortion between the practical
$\triangle_{1}$ and $\triangle_{2}$. We employ Remark \ref{rem:Region2}
to combine the sources $\left(X,Y\right)$ to a single Gaussian RV
$U$. Then we apply a polar lattice for quantization directly to the
source $U.$ Fig. \ref{fig:Simulation-performance-for-GaussianRVs}
indicates the performances of polar lattices approach the achievable
bound as the code lengths become large. Hence, the simulation results
confirm Theorem \ref{thm:Gaussian_Region2_Them} and Remark \ref{rem:Region2}.

The region $\varepsilon_{3}$ is a degenerated region. If $\frac{\delta_{2}}{\delta_{1}}<\rho^{2}$,
$R_{XY}\left(\triangle_{1},\triangle_{2}\right)=\frac{1}{2}\log\frac{1}{\triangle_{1}}$,
which coincides with the rate-distortion function of a scalar Gaussian
source. This means that the optimal coding strategy is to ignore $Y$
and simply compress $X$. Then $Y'$ can be optimally estimated from
$X'$ by $Y'=\rho X'$. The case where $\frac{\delta_{1}}{\delta_{2}}<\rho^{2}$
can be solved similarly. Therefore the construction of polar lattices
for a scalar Gaussian source given in \cite{PolarlatticeQZ} can be
applied directly for this region.

\section{Conclusion}

Explicit construction of polar codes and polar lattices for both lossy
and lossless Gray-Wyner coding is proposed. For discrete sources,
the construction of polar codes is utilized to achieve the entire
best-known region of lossless Gray-Wyner coding. The test channels
for each part of the region are identified so that the operational
meaning of the Wyner's common information can be well interpreted.
Moreover, polar codes are utilized to extract the common information
of DSBS in the lossy Gray-Wyner problem. The constructions of polar
codes to achieve the common information for each distortion region
are presented, together with the simulation result. Additionally,
an operational interpretation of the connection between the lossless
and lossy Gray-Wyner coding are given in this work. With regard to
Gaussian sources, the construction of polar lattices are shown to
be able to extract the common information for each best-known distortion
region. More importantly, it is found that the construction of a polar
lattice for extracting the common information of a pair of Gaussian
sources is equivalent to that for a single Gaussian source. This Gaussian
source can be derived simply by a linear combination of the two original
sources. Therefore, a rate-distortion bound achieving polar lattice
designed for a Gaussian source can be directly used to extract the
common information of a pair of Gaussian sources or even multiple
Gaussian sources.

\section*{Acknowledgment}

The authors would like to acknowledge Dr Naveen Goela for helpful
discussions and pointing out reference \cite{GoelaCommonInfor}.

\appendices{}

\section{Proof of Lemma \ref{lem:Lemma L Gaussian source } \label{sec:Proof-of-Lemma-LGaussian}}

For $L$ joint Gaussian RVs $\mathbf{X}_{L}=\{X_{1},X_{2},\ldots X_{L}\}$
with covariance matrix $K_{L}$ as given in (\ref{eq:covariance matrix KL}).
The determinant of the covariance matrix is 
\[
\left|K_{L}\right|=\left(1+(L-1)\rho\right)\left(1-\rho\right)^{L-1},
\]
\noindent and the inverse of $K_{L}$ is 
\[
\begin{aligned} & K_{L}^{-1}=\frac{1}{-(L-1)\rho^{2}+(L-2)\rho+1}\begin{bmatrix}(L-2)\rho+1 & -\rho & \cdots & -\rho\\
-\rho & (L-2)\rho+1 & \cdots & -\rho\\
\vdots & \vdots & \ddots & \vdots\\
-\rho & -\rho & \cdots & (L-2)\rho+1
\end{bmatrix}.\end{aligned}
\]

Therefore the joint distribution of $L$ Gaussian sources with covariance
matrix $K_{L}$ follows the next expression: {\allowdisplaybreaks

\begin{align*}
\begin{aligned} & f_{\mathbf{X}_{L}}(\mathbf{x}_{L})\\
 & =\frac{1}{\sqrt{(2\pi)^{L}\left|K_{L}\right|}}\exp\left(-\frac{1}{2}\mathbf{x}_{L}^{\mathrm{T}}K_{L}^{-1}\mathbf{x}_{L}\right)\\
 & =\frac{1}{\sqrt{(2\pi)^{L}\left(1+(L-1)\rho\right)\left(1-\rho\right)^{L-1}}}\\
 & \hspace{1em}\cdot\exp\left[-\frac{1}{2\left(-(L-1)\rho^{2}+(L-2)\rho+1\right)}\left(\left(\left(L-2\right)\rho+1\right)\sum_{i=1}^{L}x_{i}^{2}-2\rho\hspace{-2em}\sum_{\begin{aligned}i,j=1;i<j\end{aligned}
}^{L}\hspace{-2em}x_{i}x_{j}\right)\right].
\end{aligned}
\end{align*}

}

Since the components of $\mathbf{X}_{L}$ are conditionally independent
given $W$, we have {\allowdisplaybreaks[4]

\begin{equation}
\begin{aligned} & f_{\mathbf{\bar{X}}_{L}}(\mathbf{x}_{L})\\
 & =\sum_{a\in\Lambda}f_{\bar{X}_{1},\bar{X}_{2},\ldots\bar{X}_{L},\bar{W}}\left(x_{1},x_{2},\ldots x_{L},a\right)\\
 & =\sum_{a\in\Lambda}f_{\bar{W}}(a)f_{\bar{X}_{1}|\bar{W}}(x_{1}|a)f_{\bar{X}_{2}|\bar{W}}(x_{2}|a)\ldots f_{\bar{X}_{L}|\bar{W}}(x_{L}|a)\\
 & =\frac{1}{f_{\sqrt{\rho}}\left(\Lambda\right)}\sum_{a\in\Lambda}\frac{1}{\sqrt{2\pi\rho}}\exp\left(-\frac{a^{2}}{2\rho}\right)\frac{1}{\sqrt{2\pi(1-\rho)}}\exp\left(-\frac{(x_{1}-a)^{2}}{2(1-\rho)}\right)\cdots\frac{1}{\sqrt{2\pi(1-\rho)}}\exp\left(-\frac{(x_{L}-a)^{2}}{2(1-\rho)}\right)\\
 & =\frac{1}{\sqrt{(2\pi)^{L}\left(1+(L-1)\rho\right)\left(1-\rho\right)^{L-1}}}\\
 & \hspace{2em}\cdot\exp\left[-\frac{1}{2\left(-(L-1)\rho^{2}+(L-2)\rho+1\right)}\left(\left(\left(L-2\right)\rho+1\right)\sum_{i=1}^{L}x_{i}^{2}-2\rho\hspace{-2em}\sum_{\begin{aligned}i,j=1;i<j\end{aligned}
}^{L}\hspace{-2em}x_{i}x_{j}\right)\right]\\
 & \hspace{2em}\cdot\frac{1}{f_{\sqrt{\rho}}\left(\Lambda\right)}\sum_{a\in\Lambda}\frac{1}{\sqrt{\frac{2\pi\rho(1-\rho)}{1+(L-1)\rho}}}\exp\left[-\frac{1}{\frac{2\rho(1-\rho)}{1+(L-1)\rho}}\left(a-\frac{\rho}{1+(L-1)\rho}\left(x_{1}+\cdots+x_{L}\right)\right)^{2}\right]\\
 & =\frac{f_{\mathbf{X}_{L}}(\mathbf{x}_{L})}{f_{\sqrt{\rho}}\left(\Lambda\right)}\cdot\sum_{a\in\Lambda}\frac{1}{\sqrt{\frac{2\pi\rho(1-\rho)}{1+(L-1)\rho}}}\exp\left[-\frac{1}{\frac{2\rho(1-\rho)}{1+(L-1)\rho}}\left(a-\frac{\rho}{1+(L-1)\rho}\left(x_{1}+\cdots+x_{L}\right)\right)^{2}\right].
\end{aligned}
\label{eq:Gaussian joint L sources}
\end{equation}

}By the definition of the flatness factor (\ref{eq:FlatnessFactor}),
we have 
\begin{equation}
\begin{aligned} & \left|V(\Lambda)\sum_{a\in\Lambda}\frac{1}{\sqrt{\frac{2\pi\rho(1-\rho)}{1+(L-1)\rho}}}\exp\left[-\frac{1}{\frac{2\rho(1-\rho)}{1+(L-1)\rho}}\left(a-\frac{\rho\left(x_{1}+\cdots+x_{L}\right)}{1+(L-1)\rho}\right)^{2}\right]-1\right|\\
 & \leq\epsilon_{\Lambda}\left(\sqrt{\frac{\rho(1-\rho)}{1+(L-1)\rho}}\right)=\epsilon.
\end{aligned}
\label{eq:FlatnessFactorL}
\end{equation}

In this way, we derive a similar relation to (\ref{eq:GaussianFlatFactor2}).
Moreover, we have $\epsilon\left(\sqrt{\rho}\right)\leq\epsilon$
since $\epsilon_{\Lambda}\left(\sigma\right)$ is monotonically decreasing
of $\sigma$. Hence 
\begin{equation}
\left|V(\Lambda)f_{\sqrt{\rho}}(\Lambda)-1\right|\leq\epsilon\left(\sqrt{\rho}\right)\leq\epsilon.\label{eq:FlatnessFactorL2}
\end{equation}

Combining $\left(\ref{eq:Gaussian joint L sources}\right)$, $\left(\ref{eq:FlatnessFactorL}\right)$
and $\left(\ref{eq:FlatnessFactorL2}\right)$ gives us 
\[
f_{\mathbf{X}_{L}}(\mathbf{x}_{L})\left(1-2\epsilon\right)\leq f_{\mathbf{X}_{L}}(\mathbf{x}_{L})\frac{1-\epsilon}{1+\epsilon}\leq f_{\mathbf{\bar{X}}_{L}}(\mathbf{x}_{L}),
\]

\noindent and

\[
f_{\mathbf{\bar{X}}_{L}}(\mathbf{x}_{L})\leq f_{\mathbf{X}_{L}}(\mathbf{x}_{L})\frac{1+\epsilon}{1-\epsilon}\leq f_{\mathbf{X}_{L}}(\mathbf{x}_{L})\left(1+4\epsilon\right),
\]

\noindent when $\epsilon<\frac{1}{2}$. Finally, 
\[
\begin{aligned}\int_{\mathbb{R}^{L}}\left|f_{\mathbf{\bar{X}}_{L}}(\mathbf{x}_{L})-f_{\mathbf{X}_{L}}(\mathbf{x}_{L})\right|dx_{1}dx_{2}\ldots dx_{L}\leq4\epsilon\int_{\mathbb{R}^{L}}f_{\mathbf{X}_{L}}(\mathbf{x}_{L})dx_{1}dx_{2}\ldots dx_{L}=4\epsilon.\end{aligned}
\]

Similarly, the Kullback-Leibler divergence between $f_{\mathbf{\bar{X}}_{L}}(\mathbf{x}_{L})$
and $f_{\mathbf{X}_{L}}(\mathbf{x}_{L})$ can be upper-bounded as
\begin{equation}
\begin{aligned}\mathbb{D}\left(f_{\mathbf{\bar{X}}_{L}}(\mathbf{x}_{L})\|f_{\mathbf{X}_{L}}(\mathbf{x}_{L})\right) & =\int_{\mathbb{R}^{L}}f_{\mathbf{\bar{X}}_{L}}(\mathbf{x}_{L})\log\frac{f_{\mathbf{\bar{X}}_{L}}(\mathbf{x}_{L})}{f_{\mathbf{X}_{L}}(\mathbf{x}_{L})}dx_{1}dx_{2}\ldots dx_{L}\\
 & \leq\int_{\mathbb{R}^{L}}f_{\mathbf{\bar{X}}_{L}}(\mathbf{x}_{L})\log(1+4\epsilon)dx_{1}dx_{2}\ldots dx_{L}\\
 & =\log(1+4\epsilon).
\end{aligned}
\label{eq:LGaussianKLdistance}
\end{equation}

For any $\sqrt{\frac{\rho(1-\rho)}{1+(L-1)\rho}}>0$, $\epsilon$
can be made arbitrarily small by scaling $\Lambda$. Therefore, $\bar{W}$
can be viewed as the common message when $\epsilon\rightarrow0$,
according to Wyner's second approach. To show that $I\left(\mathbf{\bar{X}}_{L};\bar{W}\right)$
can be arbitrarily close to the common information, we rewrite $\mathbb{D}\left(f_{\mathbf{\bar{X}}_{L}}(\mathbf{x}_{L})\|f_{\mathbf{X}_{L}}(\mathbf{x}_{L})\right)$
as {\allowdisplaybreaks

\[
\begin{aligned} & \mathbb{D}\left(f_{\mathbf{\bar{X}}_{L}}(\mathbf{x}_{L})\|f_{\mathbf{X}_{L}}(\mathbf{x}_{L})\right)\\
 & =\int_{\mathbb{R}^{L}}f_{\mathbf{\bar{X}}_{L}}(\mathbf{x}_{L})\log\frac{f_{\mathbf{\bar{X}}_{L}}(\mathbf{x}_{L})}{f_{\mathbf{X}_{L}}(\mathbf{x}_{L})}dx_{1}dx_{2}\ldots dx_{L}\\
 & =-\int_{\mathbb{R}^{L}}f_{\mathbf{\bar{X}}_{L}}(\mathbf{x}_{L})\log f_{\mathbf{X}_{L}}(\mathbf{x}_{L})dx_{1}dx_{2}\ldots dx_{L}-h(\mathbf{\bar{X}}_{L})\\
 & =-\int_{\mathbb{R}^{L}}f_{\mathbf{\bar{X}}_{L}}(\mathbf{x}_{L})\log\left[\frac{1}{\sqrt{(2\pi)^{L}\left(1+(L-1)\rho\right)\left(1-\rho\right)^{L-1}}}\right.\\
 & \hspace{1em}\cdot\left.\exp\left[-\frac{1}{2\left(-(L-1)\rho^{2}+(L-2)\rho+1\right)}\left(\left(\left(L-2\right)\rho+1\right)\sum_{i=1}^{L}x_{i}^{2}-2\rho\hspace{-2em}\sum_{\begin{aligned}i,j=1;i<j\end{aligned}
}^{L}\hspace{-2em}x_{i}x_{j}\right)\right]\right]dx_{1}\ldots dx_{L}-h(\mathbf{\bar{X}}_{L})\\
 & =\log\left(\sqrt{(2\pi)^{L}\left(1+(L-1)\rho\right)\left(1-\rho\right)^{L-1}}\right)-h(\mathbf{\bar{X}}_{L})\\
 & \hspace{1em}+\mathsf{E_{\mathbf{\bar{X}}_{L}}}\left[\left(\left(L-2\right)\rho+1\right)\sum_{i=1}^{L}x_{i}^{2}-2\rho\hspace{-2em}\sum_{\begin{aligned}i,j=1;i<j\end{aligned}
}^{L}\hspace{-2em}x_{i}x_{j}\right]\frac{\log\left(e\right)}{2\left(-(L-1)\rho^{2}+(L-2)\rho+1\right)}\\
 & =\log\left(\sqrt{(2\pi)^{L}\left(1+(L-1)\rho\right)\left(1-\rho\right)^{L-1}}\right)-h(\mathbf{\bar{X}}_{L})+\frac{L}{2}\frac{\mathsf{E_{\bar{W}}}\left[w^{2}\right]+(L-2)\rho+1}{(L-1)\rho+1}\log\left(e\right).
\end{aligned}
\]

}Notice that $\mathsf{E}_{\bar{W}}\left[w^{2}\right]\geq\rho(1-2\epsilon)$
by \cite[Lemma 5]{LingBel13} and \cite[Remark 3]{LingBel13}. Hence{\allowdisplaybreaks

\[
\begin{aligned} & \mathbb{D}\left(f_{\mathbf{\bar{X}}_{L}}(\mathbf{x}_{L})\|f_{\mathbf{X}_{L}}(\mathbf{x}_{L})\right)\\
 & \geq\log\left(\sqrt{(2\pi)^{L}\left(1+(L-1)\rho\right)\left(1-\rho\right)^{L-1}}\right)-h(\mathbf{\bar{X}}_{L})+\left(\frac{L}{2}-\frac{L\rho}{(L-1)\rho+1}\epsilon\right)\log\left(e\right)\\
 & \geq\log\left(\sqrt{(2\pi)^{L}\left(1+(L-1)\rho\right)\left(1-\rho\right)^{L-1}}\right)-h(\mathbf{\bar{X}}_{L})+\left(\frac{L}{2}-\epsilon\right)\log\left(e\right)\\
 & =h(\mathbf{X}_{L})-h(\mathbf{\bar{X}}_{L})-\epsilon\log\left(e\right).
\end{aligned}
\]

}Using $\left(\ref{eq:LGaussianKLdistance}\right)$, we obtain 
\[
\begin{aligned}I\left(\mathbf{X}_{L};W\right)-I\left(\mathbf{\bar{X}}_{L};\bar{W}\right) & =h(\mathbf{X}_{L})-h(\mathbf{\bar{X}}_{L})\\
 & \leq\log(1+4\epsilon)+\epsilon\log\left(e\right)\\
 & \leq5\epsilon\log\left(e\right).
\end{aligned}
\]

\section{Proof of Theorem \ref{thm:Theorem L Gaussian source} \label{sec:Proof-of-Theorem-LGaussian}}

Let $\bar{W}$ be labeled by bits $\bar{W}_{1}^{r}=\{\bar{W}_{1},\ldots,\bar{W}_{r}\}$
according to a binary partition chain $\Lambda/\Lambda_{1}/\cdots/\Lambda_{r-1}/\Lambda'$
($\Lambda'$ also refers to $\Lambda_{r}$). $D_{\Lambda,\sqrt{\rho}}$
induces a distribution $P_{\bar{W}_{1}^{r}}$ whose limit corresponds
to $D_{\Lambda,\sqrt{\rho}}$ as $r\rightarrow\infty$.

By the chain rule of mutual information 
\[
I(\mathbf{\bar{X}}_{L};\bar{W}_{1}^{r})=\sum_{l=1}^{r}I(\mathbf{\bar{X}}_{L};\bar{W}_{l}|\bar{W}_{1}^{l-1}),
\]
we obtain $r$ binary-input test channels $V_{l}$ for $1\leq l\leq r$.
Given the realization $w_{1}^{l}$ of $\bar{W}_{1}^{r}$, denote by
$\mathcal{A}_{\ell}(w_{1}^{\ell})$ the coset of $\Lambda_{l}$ indexed
by $w_{1}^{l-1}$ and $w_{l}$. According to \cite{multilevel},
the channel transition PDF of the $l$-th channel $V_{l}$ is given
by {\allowdisplaybreaks

\[
\begin{aligned} & f_{\mathbf{\bar{X}}_{L}|\bar{W}_{l},\bar{W}_{1}^{l-1}}\left(\mathbf{x}_{L}|w_{l},w_{1}^{l-1}\right)\\
 & =\frac{1}{f_{\sqrt{\rho}}(\mathcal{A}_{\ell}(w_{1}^{\ell}))}\sum_{a\in\mathcal{A}_{\ell}(w_{1}^{\ell})}f_{\sqrt{\rho}}(a)f_{\mathbf{\bar{X}}_{L}|\bar{W}}(\mathbf{x}_{L}|a)\\
 & =\frac{1}{f_{\sqrt{\rho}}(\mathcal{A}_{\ell}(w_{1}^{\ell}))}\sum_{a\in\mathcal{A}_{\ell}(w_{1}^{\ell})}\frac{1}{\sqrt{2\pi\rho}}\exp\bigg(-\frac{a^{2}}{2\rho}\bigg)\frac{1}{\sqrt{2\pi(1-\rho)}}\exp\bigg(-\frac{(x_{1}-a)^{2}}{2(1-\rho)}\bigg)\cdots\frac{1}{\sqrt{2\pi(1-\rho)}}\exp\bigg(-\frac{(x_{L}-a)^{2}}{2(1-\rho)}\bigg)\\
 & =\frac{1}{\sqrt{(2\pi)^{L}\left(1+(L-1)\rho\right)\left(1-\rho\right)^{L-1}}}\\
 & \hspace{1em}\cdot\exp\left[-\frac{1}{2\left(-(L-1)\rho^{2}+(L-2)\rho+1\right)}\left(\left(\left(L-2\right)\rho+1\right)\sum_{i=1}^{L}x_{i}^{2}-2\rho\hspace{-2em}\sum_{\begin{aligned}i,j=1;i<j\end{aligned}
}^{L}\hspace{-2em}x_{i}x_{j}\right)\right]\\
 & \hspace{1em}\cdot\frac{1}{f_{\sqrt{\rho}}(\mathcal{A}_{\ell}(w_{1}^{\ell}))}\sum_{a\in\mathcal{A}_{\ell}(w_{1}^{\ell})}\frac{1}{\sqrt{\frac{2\pi\rho(1-\rho)}{1+(L-1)\rho}}}\exp\left[-\frac{1}{\frac{2\rho(1-\rho)}{1+(L-1)\rho}}\left(a-\frac{\rho}{1+(L-1)\rho}\left(x_{1}+\cdots+x_{L}\right)\right)^{2}\right],
\end{aligned}
\]

}

Let $\tilde{V}_{l}$ be a symmetrized channel with input $\tilde{W}_{l}$
(assume to be uniformly distributed) and output $\left(\mathbf{\bar{X}}_{L},\bar{W}_{1}^{\ell-1},\bar{W}_{l}\oplus\tilde{W}_{l}\right)$,
built from the asymmetric channel $V_{l}$. Then the joint PDF of
$V_{l}$ can be represented by the transition PDF of $\tilde{V}_{l}$
(see \cite{polarlatticeJ} for more details), as shown in the following
equation. {\allowdisplaybreaks

\begin{equation}
\begin{aligned} & f_{\tilde{V}_{\ell}}(\mathbf{x}_{L},w_{1}^{\ell-1},w_{\ell}\oplus\tilde{w}_{\ell}|\tilde{w}_{\ell})\\
 & =f_{\mathbf{\bar{X}}_{L},\bar{W}_{1}^{\ell}}(\mathbf{x}_{L},w_{1}^{\ell})\\
 & =\frac{1}{\sqrt{(2\pi)^{L}\left(1+(L-1)\rho\right)\left(1-\rho\right)^{L-1}}}\\
 & \hspace{1em}\cdot\exp\left[-\frac{1}{2\left(-(L-1)\rho^{2}+(L-2)\rho+1\right)}\left(\left(\left(L-2\right)\rho+1\right)\sum_{i=1}^{L}x_{i}^{2}-2\rho\hspace{-2em}\sum_{\begin{aligned}i,j=1;i<j\end{aligned}
}^{L}\hspace{-2em}x_{i}x_{j}\right)\right]\\
 & \hspace{1em}\cdot\frac{1}{f_{\sqrt{\rho}}(\Lambda)}\sum_{a\in\mathcal{A}_{\ell}(w_{1}^{\ell})}\frac{1}{\sqrt{\frac{2\pi\rho(1-\rho)}{1+(L-1)\rho}}}\exp\left[-\frac{1}{\frac{2\rho(1-\rho)}{1+(L-1)\rho}}\left(a-\frac{L\rho}{1+(L-1)\rho}\frac{\left(x_{1}+\cdots+x_{L}\right)}{L}\right)^{2}\right].
\end{aligned}
\label{eq:AppTheoremSymm}
\end{equation}

}

Comparing with the $\Lambda_{\ell-1}/\Lambda_{\ell}$ channel \cite[Equation (13)]{polarlatticeJ},
we see that the symmetrized channel (\ref{eq:GaussianSymmetrimizedF})
is equivalent to a $\Lambda_{\ell-1}/\Lambda_{\ell}$ channel with
noise variance $\frac{\rho(1-\rho)}{1+(L-1)\rho}$ in the sense of
the likelihood ratio.

$\mathbf{X}_{L}=\{X_{1},X_{2},\ldots X_{L}\}$ are Gaussian RVs with
zero mean and covariance matrix $K_{L}$ as given in (\ref{eq:covariance matrix KL}).
The mean value and variance of the Gaussian RV $\frac{X_{1}+\cdots+X_{L}}{L}$
are respectively 
\[
\mathsf{E}\left[\frac{X_{1}+\cdots+X_{L}}{L}\right]=\frac{1}{L}\left(\mathsf{E}\left[X_{1}\right]+\ldots+\mathsf{E}\left[X_{L}\right]\right)=0,
\]
and 

\[
\begin{aligned}\sigma^{2}\left(\frac{X_{1}+\cdots+X_{L}}{L}\right) & \mathsf{=E}\left[\left(\frac{X_{1}+\cdots+X_{L}}{L}\right)^{2}\right]-\left(\mathsf{E}\left[\frac{X_{1}+\cdots+X_{L}}{L}\right]\right)^{2}\\
 & =\frac{1}{L^{2}}\left(\mathsf{E}\left[\sum_{i=1}^{L}X_{i}^{2}+2\rho\hspace{-2em}\sum_{\begin{aligned}i,j=1;i<j\end{aligned}
}^{L}\hspace{-2em}X_{i}X_{j}\right]\right)\\
 & =\frac{L+2\rho\begin{pmatrix}L\\
2
\end{pmatrix}}{L^{2}}=\frac{1+\left(L-1\right)\rho}{L},
\end{aligned}
\]
where $\begin{pmatrix}L\\
r
\end{pmatrix}\triangleq\frac{L!}{r!\left(L-r\right)!}$. 

Consider the construction of a polar lattice to quantize $\frac{X_{1}+X_{2}+\ldots+X_{L}}{L}$
using reconstruction distribution $D_{\Lambda,\sqrt{\rho}}$. Denote
the variance of the source and the reconstruction by $\sigma_{s}^{2}=\frac{1+\left(L-1\right)\rho}{L}$
and $\sigma_{r}^{2}=\rho$, respectively. Thus, the variance of the
noise equals $\sigma_{z}^{2}=\sigma_{s}^{2}-\sigma_{r}^{2}=\frac{1-\rho}{L}$.
Then we apply MMSE to this relation. By definitions, the MMSE scaling
coefficient $\alpha$ and noise variance $\tilde{\sigma}_{z}^{2}$
are given by 
\[
\alpha=\frac{\sigma_{r}^{2}}{\sigma_{s}^{2}}=\frac{L\rho}{1+\left(L-1\right)\rho},
\]

\[
\tilde{\sigma}_{z}^{2}=\alpha\cdot\sigma_{z}^{2}=\frac{\rho(1-\rho)}{1+\left(L-1\right)\rho},
\]
which are the same to those in the summation section of $\left(\ref{eq:AppTheoremSymm}\right)$.

\section{Proof of Lemma \ref{lem:TwoGaussianRegion2} \label{sec:Proof-of-Lemma-TwoGaussianRegion2}}

The PDF of $\left(X,Y\right)$ can be represented from the PDF of
$\left(\bar{X},\bar{Y}\right)$ as follows. {\allowdisplaybreaks

\begin{equation}
\begin{aligned} & f_{\bar{X},\bar{Y}}\left(x,y\right)\\
 & =\sum_{a\in\varLambda}f_{\bar{X'}}(a)f_{\bar{X},\bar{Y}|\bar{X'}}\left(x,y|a\right)\\
 & =\sum_{a\in\varLambda}f_{\bar{X'}}(a)f_{Z_{1}Z_{2}}\left(x-a,y-\sqrt{\frac{\delta_{2}}{\delta_{1}}}a\right)\\
 & =\frac{1}{f_{\sqrt{\delta_{1}}}\left(\Lambda\right)}\sum_{a\in\Lambda}\frac{1}{\sqrt{2\pi\delta_{1}}}\exp\left(-\frac{a^{2}}{2\delta_{1}}\right)\frac{1}{2\pi\sqrt{\triangle_{1}\triangle_{2}-\left(\rho-\sqrt{\delta_{1}\delta_{2}}\right)^{2}}}\\
 & \hspace{1em}\cdot\exp\left[-\frac{1}{2\left(1-\frac{\left(\rho-\sqrt{\delta_{1}\delta_{2}}\right)^{2}}{\triangle_{1}\triangle_{2}}\right)}\left(\frac{\left(x-a\right)^{2}}{\triangle_{1}}+\frac{\left(y-\sqrt{\frac{\delta_{2}}{\delta_{1}}}a\right)^{2}}{\triangle_{2}}-\frac{2\left(\rho-\sqrt{\delta_{1}\delta_{2}}\right)\left(x-a\right)\left(y-\sqrt{\frac{\delta_{2}}{\delta_{1}}}a\right)}{\triangle_{1}\triangle_{2}}\right)\right]\\
 & =\frac{1}{2\pi\sqrt{1-\rho^{2}}}\exp\left[-\frac{\left(x^{2}+y^{2}-2\rho xy\right)}{2\left(1-\rho^{2}\right)}\right]\frac{1}{f_{\sqrt{\delta_{1}}}\left(\Lambda\right)}\sum_{a\in\Lambda}\sqrt{\frac{1-\rho^{2}}{2\pi\delta_{1}\left(\triangle_{1}\triangle_{2}-\left(\rho-\sqrt{\delta_{1}\delta_{2}}\right)^{2}\right)}}\\
 & \hspace{1em}\cdot\exp\left[-\frac{1-\rho^{2}}{2\delta_{1}\left(\triangle_{1}\triangle_{2}-\left(\rho-\sqrt{\delta_{1}\delta_{2}}\right)^{2}\right)}\left(a-\frac{\left(\delta_{1}-\rho\sqrt{\delta_{1}\delta_{2}}\right)x+\left(\sqrt{\delta_{1}\delta_{2}}-\delta_{1}\rho\right)y}{1-\rho^{2}}\right)^{2}\right],
\end{aligned}
\label{eq:Region2GaussianJointProb}
\end{equation}
}where $\frac{1}{2\pi\sqrt{1-\rho^2}} \exp \bigg( -\frac{x^2+y^2-2\rho xy}{2(1-\rho^2)} \bigg)=f_{X,Y}(x,y)$
is the PDF of two joint Gaussian RVs. By the definition of the flatness
factor (\ref{eq:FlatnessFactor}), we have

{\allowdisplaybreaks
\begin{equation}
\begin{aligned} & \left|V\left(\Lambda\right)\sum_{a\in\Lambda}\sqrt{\frac{1-\rho^{2}}{2\pi\delta_{1}\left(\triangle_{1}\triangle_{2}-\left(\rho-\sqrt{\delta_{1}\delta_{2}}\right)^{2}\right)}}\right.\\
 & \left.\hspace{1em}\cdot\exp\left[-\frac{1-\rho^{2}}{2\delta_{1}\left(\triangle_{1}\triangle_{2}-\left(\rho-\sqrt{\delta_{1}\delta_{2}}\right)^{2}\right)}\left(a-\frac{\left(\delta_{1}-\rho\sqrt{\delta_{1}\delta_{2}}\right)x+\left(\sqrt{\delta_{1}\delta_{2}}-\delta_{1}\rho\right)y}{1-\rho^{2}}\right)^{2}\right]-1\right|\\
 & \leq\epsilon_{\Lambda}\left(\sqrt{\frac{\delta_{1}\left(\triangle_{1}\triangle_{2}-\left(\rho-\sqrt{\delta_{1}\delta_{2}}\right)^{2}\right)}{1-\rho^{2}}}\right)=\epsilon.
\end{aligned}
\label{eq:Region2GaussianFlatFactor}
\end{equation}

}

Since $\epsilon_{\Lambda}\left(\sigma\right)$ is a monotonically
decreasing function of $\sigma$ and the fact that 
\[
\begin{aligned} & \triangle_{1}\triangle_{2}-\left(\rho-\sqrt{\delta_{1}\delta_{2}}\right)^{2}-1+\rho^{2}\\
 & =2\rho\sqrt{\delta_{1}\delta_{2}}-\delta_{1}-\delta_{2}\\
 & =-\left(\sqrt{\delta_{1}}-\sqrt{\delta_{2}}\right)^{2}-2\sqrt{\delta_{1}\delta_{2}}\left(1-\rho\right)\leq0,
\end{aligned}
\]
we have

\[
\begin{aligned}\frac{\triangle_{1}\triangle_{2}-\left(\rho-\sqrt{\delta_{1}\delta_{2}}\right)^{2}}{1-\rho^{2}} & \leq1.\end{aligned}
\]
Therefore, it implies $\epsilon\left(\sqrt{\delta_{1}}\right)\leq\epsilon$
and more specifically 
\begin{equation}
\left|V\left(\Lambda\right)f_{\sqrt{\delta_{1}}}\left(\Lambda\right)-1\right|\leq\epsilon.\label{eq:Region2GaussianFlatFactor2}
\end{equation}

From the above results (\ref{eq:Region2GaussianJointProb}), (\ref{eq:Region2GaussianFlatFactor})
and (\ref{eq:Region2GaussianFlatFactor2}), we have 
\[
f_{X,Y}\left(x,y\right)\left(1-2\epsilon\right)\leq f_{X,Y}\left(x,y\right)\frac{1-\epsilon}{1+\epsilon}\leq f_{\bar{X},\bar{Y}}\left(x,y\right),
\]
and 
\[
f_{\bar{X},\bar{Y}}\left(x,y\right)\leq f_{X,Y}\left(x,y\right)\frac{1+\epsilon}{1-\epsilon}\leq f_{X,Y}\left(x,y\right)\left(1-4\epsilon\right),
\]
when $\epsilon<0.5$. Finally, 

\begin{align*}
\int_{\mathbb{R}^{2}}\left|f_{\bar{X},\bar{Y}}(x,y)-f_{X,Y}(x,y)\right|dxdy\\
\leq4\epsilon\int_{\mathbb{R}^{2}}f_{X,Y}(x,y)dxdy=4\epsilon.
\end{align*}

Similarly, the Kullback-Leibler divergence between $f_{\bar{X},\bar{Y}}(x,y)$
and $f_{X,Y}(x,y)$ can be upper-bounded as 
\begin{equation}
\begin{aligned}\mathbb{D}(f_{\bar{X},\bar{Y}}\|f_{X,Y}) & =\int_{\mathbb{R}^{2}}f_{\bar{X},\bar{Y}}(x,y)\log\frac{f_{\bar{X},\bar{Y}}(x,y)}{f_{X,Y}(x,y)}dxdy\\
 & \leq\int_{\mathbb{R}^{2}}f_{\bar{X},\bar{Y}}(x,y)\log(1+4\epsilon)dxdy\\
 & =\log(1+4\epsilon).
\end{aligned}
\label{eq:GaussianDistanceRegion2}
\end{equation}

For any $\sqrt{\frac{\delta_{1}\left(\triangle_{1}\triangle_{2}-\left(\rho-\sqrt{\delta_{1}\delta_{2}}\right)^{2}\right)}{1-\rho^{2}}}>0$,
$\epsilon$ can be made arbitrarily small by scaling $\Lambda$. To
show that $I\left(\bar{X},\bar{Y};\bar{X'}\right)$ can be arbitrarily
close to $I\left(X,Y;X'\right)$, we rewrite $\mathbb{D}(f_{\bar{X},\bar{Y}}\|f_{X,Y})$
as

\[
\begin{aligned} & \mathbb{D}(f_{\bar{X},\bar{Y}}\|f_{X,Y})\\
 & =\int_{\mathbb{R}^{2}}f_{\bar{X},\bar{Y}}(x,y)\log\frac{f_{\bar{X},\bar{Y}}(x,y)}{f_{X,Y}(x,y)}dxdy\\
 & =-\int_{\mathbb{R}^{2}}f_{\bar{X},\bar{Y}}(x,y)\log f_{X,Y}(x,y)dxdy-h(\bar{X},\bar{Y})\\
 & =-\int_{\mathbb{R}^{2}}f_{\bar{X},\bar{Y}}(x,y)\log\left(\frac{1}{2\pi\sqrt{1-\rho^{2}}}\exp\bigg(-\frac{x^{2}+y^{2}-2\rho xy}{2(1-\rho^{2})}\bigg)\right)dxdy-h(\bar{X},\bar{Y})\\
 & =\log\big(2\pi\sqrt{1-\rho^{2}}\big)+\frac{\mathsf{E}_{\bar{X},\bar{Y}}[x^{2}+y^{2}-2\rho xy]}{2(1-\rho^{2})}\log(e)-h(\bar{X},\bar{Y})\\
 & =\log\big(2\pi\sqrt{1-\rho^{2}}\big)+\log(e)-h(\bar{X},\bar{Y}),
\end{aligned}
\]
based on the fact that $\mathsf{E}_{\bar{X},\bar{Y}}\left[x,y\right]=\rho$
and $\mathsf{E}_{\bar{X},\bar{Y}}\left[x^{2}\right]=\mathsf{E}_{\bar{X},\bar{Y}}\left[y^{2}\right]=1$.
Trivially we have 
\[
\begin{aligned}\mathbb{D}(f_{\bar{X},\bar{Y}}\|f_{X,Y}) & \geq\log\big(2\pi\sqrt{1-\rho^{2}}\big)+(1-\epsilon)\log(e)-h(\bar{X},\bar{Y})\\
 & =h(X,Y)-h(\bar{X},\bar{Y})-\epsilon\log(e).
\end{aligned}
\]

Using (\ref{eq:GaussianDistanceRegion2}), we obtain 
\[
\begin{aligned}I(X,Y;X')-I(\bar{X},\bar{Y};\bar{X'}) & =h(X,Y)-h(\bar{X},\bar{Y})\\
 & \leq\log(1+4\epsilon)+\epsilon\log(e)\\
 & \leq5\epsilon\log(e).
\end{aligned}
\]

\section{Proof of Theorem \ref{thm:Gaussian_Region2_Them} \label{sec:Proof-of-Theorem-Achivability}}

Since this proof uses multilevel coding, the notations are changed
differently from the rest of the paper. 

\textit{Notations:} Denote $X_{l}$ a RV $X$ at level $l$. The $i$-th
realization of $X_{l}$ is denoted by $x_{l}^{i}$. The notation $x_{l}^{i:j}$
denotes vector $\left(x_{l}^{i},\ldots x_{l}^{j}\right)$, which is
a realization of RVs $X_{l}^{i:j}=\left(X_{l}^{i},\ldots X_{l}^{j}\right)$.
Similarly, $x_{l:j}^{i}$ denotes the realization of the $i$-th RV
from level $l$ to level $j$. $X_{l}^{\mathcal{I}}$ denotes the
subvector $\left\{ X_{l}^{i}\right\} _{i\in\mathcal{I}}$ at level
$l$. For the consistency of notations in this proof, let $X^{1:N}$
denote a vector $\left(X^{1},\ldots,X^{N}\right)$ and $x^{i}$ denote
the $i$-th realization of RV $X^{i}$. 

Firstly, for the sources $\left(\bar{X},\bar{Y}\right)$ and reconstruction
RVs $\left(\bar{X'},\bar{Y'}\right)$, we consider the average performance
of the multilevel polar codes with all possible choice of frozen sets
$u_{l}^{\mathcal{F}_{l}}$ (defined in \cite[Equation (16)]{PolarlatticeQZ})
at each level. If the encoding rule described in the form of \cite[Equation (17)]{PolarlatticeQZ}
is used for all $i\in[N]$ at each level, the resulted average distortions
of $\left(\bar{X},\bar{Y}\right)$ are given by 
\[
\begin{aligned}\triangle_{P,\bar{X}} & =\frac{1}{N}\sum_{u_{1:r}^{1:N},\bar{x}^{1:N}}P_{U_{1:r}^{1:N},\bar{X}^{1:N}}\left(u_{1:r}^{1:N},\bar{x}^{1:N}\right)d\left(\bar{x}^{1:N},\mathcal{M}\left(u_{1:r}^{1:N}G_{N}\right)\right),\\
\triangle_{P,\bar{Y}} & =\frac{1}{N}\sum_{u_{1:r}^{1:N},\bar{y}^{1:N}}P_{U_{1:r}^{1:N},\bar{Y}^{1:N}}\left(u_{1:r}^{1:N},\bar{y}^{1:N}\right)d\left(\bar{y}^{1:N},\mathcal{M}\left(\sqrt{\frac{\delta_{2}}{\delta_{1}}}u_{1:r}^{1:N}G_{N}\right)\right),
\end{aligned}
\]
 where $\mathcal{M}\left(u_{1:r}^{1:N}G_{N}\right)$ denotes a mapping
from $u_{1:r}^{1:N}$ to $\bar{x'}{}^{1:N}$ according to \cite[Equation (38)]{PolarlatticeQZ}.
Due to the linear relation $\bar{Y'}=\sqrt{\frac{\delta_{2}}{\delta_{1}}}\bar{X'}$
for region $\varepsilon_{2}$, we have $\bar{x'}{}_{l}^{1:N}=u_{l}^{1:N}G_{N}$
and $\bar{y'}{}_{l}^{1:N}=\sqrt{\frac{\delta_{2}}{\delta_{1}}}u_{l}^{1:N}G_{N}$
for each level. When $r\rightarrow\infty$, there exist an one-to-one
mapping from $u_{1:r}^{1:N}$ to $\bar{x'}{}^{1:N}$ and $\bar{y'}{}^{1:N}$.
Thus, we have {\allowdisplaybreaks 
\[
\begin{aligned} & \triangle_{P,\bar{X}}\\
 & =\frac{1}{N}\sum_{\bar{x'}^{1:N},\bar{x}^{1:N}}P_{\bar{X'}^{1:N},\bar{X}^{1:N}}\left(\bar{x'}^{1:N},\bar{x}^{1:N}\right)d^{N}\left(\bar{x'}^{1:N},\bar{x}^{1:N}\right)\\
 & =N\cdot\frac{1}{N}\sum_{\bar{x'},\bar{x}}P_{\bar{X'},\bar{X}}\left(\bar{x'},\bar{x}\right)d\left(\bar{x'},\bar{x}\right)\\
 & =\sum_{\bar{x'}\in\Lambda}P_{\bar{X'}}\left(\bar{x'}\right)\int_{-\infty}^{+\infty}\frac{1}{\sqrt{2\pi\triangle_{1}}}\exp\left(-\frac{\left(\bar{x}-\bar{x'}\right)^{2}}{2\triangle_{1}}\right)\left(\bar{x}-\bar{x'}\right)^{2}d\bar{x}\\
 & =\triangle_{1}.
\end{aligned}
\]

}

We can apply this distortion to source $\bar{Y}$ in a similar manner
and derive $\triangle_{P,\bar{Y}}=\triangle_{2}$. The results $\triangle_{P,\bar{X}}=\triangle_{1}$
and $\triangle_{P,\bar{Y}}=\triangle_{2}$ are reasonable since the
encoder does not do any compression. Next we replace $P_{U_{1:r}^{1:N},\bar{X}^{1:N}}\left(u_{1:r}^{1:N},\bar{x}^{1:N}\right)$
to $Q_{U_{1:r}^{1:N},\bar{X}^{1:N}}\left(u_{1:r}^{1:N},\bar{x}^{1:N}\right)$
and $P_{U_{1:r}^{1:N},\bar{Y}^{1:N}}\left(u_{1:r}^{1:N},\bar{y}^{1:N}\right)$
to $Q_{U_{1:r}^{1:N},\bar{Y}^{1:N}}\left(u_{1:r}^{1:N},\bar{y}^{1:N}\right)$,
so that the encoder compresses $\left(\bar{X}^{1:N},\bar{Y}^{1:N}\right)$
to $U_{l}^{\mathcal{I}_{l}}$ at each level according to the rule
\cite[Equation (17)]{PolarlatticeQZ}. The result average distortion
$\triangle_{Q,\bar{X}}$ can be bounded as {\allowdisplaybreaks 
\[
\begin{aligned} & \triangle_{Q,\bar{X}}\\
 & =\frac{1}{N}\sum_{u_{1:r}^{1:N},\bar{x}^{1:N}}Q_{U_{1:r}^{1:N},\bar{X}^{1:N}}\left(u_{1:r}^{1:N},\bar{x}^{1:N}\right)d\left(\bar{x}^{1:N},\mathcal{M}\left(u_{1:r}^{1:N}G_{N}\right)\right)\\
 & \leq\frac{1}{N}\sum_{u_{1:r}^{1:N},\bar{x}^{1:N}}P_{U_{1:r}^{1:N},\bar{X}^{1:N}}\left(u_{1:r}^{1:N},\bar{x}^{1:N}\right)d\left(\bar{x}^{1:N},\mathcal{M}\left(u_{1:r}^{1:N}G_{N}\right)\right)\\
 & \hspace{1em}+\frac{1}{N}\sum_{u_{1:r}^{1:N},\bar{x}^{1:N}}\left|P_{U_{1:r}^{1:N},\bar{X}^{1:N}}\left(u_{1:r}^{1:N},\bar{x}^{1:N}\right)-Q_{U_{1:r}^{1:N},\bar{X}^{1:N}}\left(u_{1:r}^{1:N},\bar{x}^{1:N}\right)\right|d\left(\bar{x}^{1:N},\mathcal{M}\left(u_{1:r}^{1:N}G_{N}\right)\right)\\
 & \leq\triangle_{P,\bar{X}}+\frac{1}{N}\cdot Nd_{\max x}2\mathbb{V}\left(P_{U_{1:r}^{1:N},\bar{X}^{1:N}}\left(u_{1:r}^{1:N},\bar{x}^{1:N}\right),Q_{U_{1:r}^{1:N},\bar{X}^{1:N}}\left(u_{1:r}^{1:N},\bar{x}^{1:N}\right)\right)\\
 & =\triangle_{1}+\mathcal{O}\left(2^{-N^{\beta'}}\right),
\end{aligned}
\]
}where $d_{\max x}$ is assumed to be the maximum distortion between
$\bar{x}$ and $\bar{x'}$. The last equality follows from \cite[Equation (20)]{PolarlatticeQZ}
and $r=\mathcal{O}\left(\log\log N\right)$ \cite[Lemma 5]{polarlatticeJ}.
Similarly, we also have $\triangle_{Q,\bar{Y}}=\triangle_{2}+\mathcal{O}\left(2^{-N^{\beta'}}\right)$
for source $\bar{Y}$.

Now we quantize the Gaussian sources $\left(X,Y\right)$ by the same
encoder. Again we take the source $X$ as example. The resulted distortion
$\triangle_{Q,X}$ can be written as {\allowdisplaybreaks 
\[
\begin{aligned} & \triangle_{Q,X}\\
 & =\frac{1}{N}\sum_{u_{1:r}^{1:N},x^{1:N}}Q_{U_{1:r}^{1:N},X^{1:N}}\left(u_{1:r}^{1:N},x^{1:N}\right)d\left(x^{1:N},\mathcal{M}\left(u_{1:r}^{1:N}G_{N}\right)\right)\\
 & =\frac{1}{N}\sum_{u_{1:r}^{1:N},x^{1:N}}P_{X^{1:N}}\left(x^{1:N}\right)Q_{U_{1:r}^{1:N}\mid X^{1:N}}\left(u_{1:r}^{1:N}\mid x^{1:N}\right)d\left(x^{1:N},\mathcal{M}\left(u_{1:r}^{1:N}G_{N}\right)\right),
\end{aligned}
\]

}

Since the same encoder is used, we apply the same realizations $\left(x^{1:N},y^{1:N}\right)$
for both RV pairs $\left(X^{1:N},Y^{1:N}\right)$ and $\left(\bar{X}^{1:N},\bar{Y}^{1:N}\right)$.
Then the relation holds 
\[
\begin{aligned}Q_{U_{1:r}^{1:N}\mid X^{1:N}}\left(u_{1:r}^{1:N}\mid x^{1:N}\right) & =Q_{U_{1:r}^{1:N}\mid\bar{X}^{1:N}}\left(u_{1:r}^{1:N}\mid x^{1:N}\right),\end{aligned}
\]
and hence {\allowdisplaybreaks 
\[
\begin{aligned} & \triangle_{Q,X}-\triangle_{Q,\bar{X}}\\
 & =\frac{1}{N}\sum_{u_{1:r}^{1:N},x^{1:N}}\left(P_{X^{1:N}}\left(x^{1:N}\right)-P_{\bar{X}^{1:N}}\left(x^{1:N}\right)\right)\\
 & \hspace{2em}\cdot Q_{U_{1:r}^{1:N}\mid X^{1:N}}\left(u_{1:r}^{1:N}\mid x^{1:N}\right)\cdot d\left(x^{1:N},\mathcal{M}\left(u_{1:r}^{1:N}G_{N}\right)\right)\\
 & \leq\frac{1}{N}\sum_{x^{1:N}}\left|P_{X^{1:N}}\left(x^{1:N}\right)-P_{\bar{X}^{1:N}}\left(x^{1:N}\right)\right|Nd_{\max x}.
\end{aligned}
\]

}

By the telescoping expansion, {\allowdisplaybreaks 
\[
\begin{aligned} & \sum_{x^{1:N}}\left|P_{X^{1:N}}\left(x^{1:N}\right)-P_{\bar{X}^{1:N}}\left(x^{1:N}\right)\right|\\
 & =\sum_{x^{1:N}}\sum_{i=1}^{N}\left|P_{X^{i}}\left(x^{i}\right)-P_{\bar{X}^{i}}\left(x^{i}\right)\right|P_{X^{1:i-1}}\left(x^{1:i-1}\right)P_{\bar{X}^{i+1:N}}\left(x^{i+1:N}\right)\\
 & =\sum_{i=1}^{N}\sum_{x^{i}}\left|P_{X^{i}}\left(x^{i}\right)-P_{\bar{X}^{i}}\left(x^{i}\right)\right|\\
 & =\sum_{i=1}^{N}\sum_{x^{i}}\left|\sum_{y^{i}}\left(P_{X^{i},Y^{i}}\left(x^{i},y^{i}\right)-P_{\bar{X}^{i},\bar{Y}^{i}}\left(x^{i},y^{i}\right)\right)\right|\\
 & \leq\sum_{i=1}^{N}\sum_{x^{i}}\sum_{y^{i}}\left|P_{X^{i},Y^{i}}\left(x^{i},y^{i}\right)-P_{\bar{X}^{i},\bar{Y}^{i}}\left(x^{i},y^{i}\right)\right|\\
 & \leq N\cdot4\epsilon_{\Lambda}\left(\sqrt{\frac{\delta_{1}\left(\triangle_{1}\triangle_{2}-\left(\rho-\sqrt{\delta_{1}\delta_{2}}\right)^{2}\right)}{1-\rho^{2}}}\right).
\end{aligned}
\]
}The last inequality results from Lemma \ref{lem:TwoGaussianRegion2}.

As a result, 
\begin{equation}
\begin{aligned}\triangle_{Q,X} & \leq\triangle_{1}+\mathcal{O}\left(2^{-N^{\beta'}}\right)+N\cdot4\epsilon_{\Lambda}\left(\sqrt{\frac{\delta_{1}\left(\triangle_{1}\triangle_{2}-\left(\rho-\sqrt{\delta_{1}\delta_{2}}\right)^{2}\right)}{1-\rho^{2}}}\right)d_{\max x}.\end{aligned}
\label{eq:Gaussian_Distortion_Conditions}
\end{equation}
Similarly, the distortion of $Y$ can be bounded as 
\begin{equation}
\triangle_{Q,Y}\leq\triangle_{2}+\mathcal{O}\left(2^{-N^{\beta'}}\right)+N\cdot4\epsilon_{\Lambda}\left(\sqrt{\frac{\delta_{1}\left(\triangle_{1}\triangle_{2}-\left(\rho-\sqrt{\delta_{1}\delta_{2}}\right)^{2}\right)}{1-\rho^{2}}}\right)d_{\max y},\label{eq:Gaussian_Distortion_Conditions_Y}
\end{equation}
where $d_{\max y}$ is assumed to be the maximum distortion between
$\bar{y}$ and $\bar{y'}$.

By scaling $\Lambda$, we can make 
\[
\epsilon_{\Lambda}\left(\sqrt{\frac{\delta_{1}\left(\triangle_{1}\triangle_{2}-\left(\rho-\sqrt{\delta_{1}\delta_{2}}\right)^{2}\right)}{1-\rho^{2}}}\right)\ll\frac{1}{4N\cdot\max\left(d_{\max x},d_{\max y}\right)}.
\]
Therefore, $\triangle_{Q,X}$ and $\triangle_{Q,Y}$ can be arbitrarily
close to $\triangle_{1}$ and $\triangle_{2}$, respectively, with
the rate 
\[
\begin{aligned}R & >I\left(\bar{X},\bar{Y};\bar{X'}\right)\\
 & \geq\frac{1}{2}\log\frac{1-\rho^{2}}{\triangle_{1}\triangle_{2}-\left(\rho-\sqrt{\left(1-\triangle_{1}\right)\left(1-\triangle_{2}\right)}\right)^{2}}-5\epsilon_{\Lambda}\left(\sqrt{\frac{\delta_{1}\left(\triangle_{1}\triangle_{2}-\left(\rho-\sqrt{\delta_{1}\delta_{2}}\right)^{2}\right)}{1-\rho^{2}}}\right)\log\left(e\right).
\end{aligned}
\]
When $\epsilon_{\Lambda}\left(\sqrt{\frac{\delta_{1}\left(\triangle_{1}\triangle_{2}-\left(\rho-\sqrt{\delta_{1}\delta_{2}}\right)^{2}\right)}{1-\rho^{2}}}\right)\rightarrow0$,
we have 
\[
I\left(\bar{X},\bar{Y};\bar{X'}\right)\rightarrow\frac{1}{2}\log\frac{1-\rho^{2}}{\triangle_{1}\triangle_{2}-\left(\rho-\sqrt{\left(1-\triangle_{1}\right)\left(1-\triangle_{2}\right)}\right)^{2}}
\]
 and 
\[
R>\frac{1}{2}\log\frac{1-\rho^{2}}{\triangle_{1}\triangle_{2}-\left(\rho-\sqrt{\left(1-\triangle_{1}\right)\left(1-\triangle_{2}\right)}\right)^{2}}.
\]

Since $\triangle_{Q,X}$ and $\triangle_{Q,Y}$ are average distortions
over all random choices of $u_{l}^{\mathcal{F}_{l}}$, there exists
at least one specific choice of $u_{l}^{\mathcal{F}_{l}}$ at each
level making the average distortions satisfying (\ref{eq:Gaussian_Distortion_Conditions})
and (\ref{eq:Gaussian_Distortion_Conditions_Y}). This is a shift
on the constructed polar lattice. As a result, the shifted polar lattice
achieves the rate-distortion bound of the Gaussian sources.

\bibliographystyle{IEEEtran}
\bibliography{Myreff}

\begin{thebibliography}{10}
\providecommand{\url}[1]{#1}
\csname url@samestyle\endcsname
\providecommand{\newblock}{\relax}
\providecommand{\bibinfo}[2]{#2}
\providecommand{\BIBentrySTDinterwordspacing}{\spaceskip=0pt\relax}
\providecommand{\BIBentryALTinterwordstretchfactor}{4}
\providecommand{\BIBentryALTinterwordspacing}{\spaceskip=\fontdimen2\font plus
\BIBentryALTinterwordstretchfactor\fontdimen3\font minus
  \fontdimen4\font\relax}
\providecommand{\BIBforeignlanguage}[2]{{%
\expandafter\ifx\csname l@#1\endcsname\relax
\typeout{** WARNING: IEEEtran.bst: No hyphenation pattern has been}%
\typeout{** loaded for the language `#1'. Using the pattern for}%
\typeout{** the default language instead.}%
\else
\language=\csname l@#1\endcsname
\fi
#2}}
\providecommand{\BIBdecl}{\relax}
\BIBdecl

\bibitem{GeXulossyCI}
G.~Xu, W.~Liu, and B.~Chen, ``A lossy source coding interpretation of
  {W}yner{'}s common information,'' \emph{IEEE Trans. Inf. Theory}, vol.~62,
  pp. 754 -- 768, 2016.

\bibitem{shannon2001mathematical}
C.~Shannon, ``A mathematical theory of communication,'' \emph{Bell Syst. Tech.
  J.}, vol.~27, no.~3, pp. 379--423, July 1948.

\bibitem{gacs1973common}
P.~G{\'a}cs and J.~K{\"o}rner, ``Common information is far less than mutual
  information,'' \emph{Problems Contr. Inform. Theory}, vol.~2, no.~2, pp.
  149--162, 1973.

\bibitem{WynerCI}
A.~D. Wyner, ``The common information of two dependent random variables,''
  \emph{IEEE Trans. Inf. Theory}, vol.~21, no.~2, pp. 163--179, March 1975.

\bibitem{Ahlswede1993Part1}
R.~Ahlswede and I.~Csiszar, ``Common randomness in information theory and
  cryptography. {I}. secret sharing,'' \emph{IEEE Trans. Inf. Theory}, vol.~39,
  no.~4, pp. 1121--1132, Jul 1993.

\bibitem{Ahlswede1998Part2}
------, ``Common randomness in information theory and cryptography. {II}. {CR}
  capacity,'' \emph{IEEE Trans. Inf. Theory}, vol.~44, no.~1, pp. 225--240, Jan
  1998.

\bibitem{Maurer1993Secret}
U.~M. Maurer, ``Secret key agreement by public discussion from common
  information,'' \emph{IEEE Trans. Inf. Theory}, vol.~39, no.~3, pp. 733--742,
  May 1993.

\bibitem{Viswanatha2011Gray}
K.~Viswanatha, E.~Akyol, and K.~Rose, ``An optimal transmit-receive rate
  tradeoff in {G}ray-{W}yner network and its relation to common information,''
  in \emph{Proc. 2011 IEEE Inform. Theory Workshop}, Oct 2011, pp. 105--109.

\bibitem{gray1974source}
R.~Gray and A.~Wyner, ``Source coding for a simple network,'' \emph{Bell System
  Technical Journal}, vol.~53, no.~9, pp. 1681--1721, 1974.

\bibitem{viswanatha2014lossy}
K.~B. Viswanatha, E.~Akyol, and K.~Rose, ``The lossy common information of
  correlated sources,'' \emph{IEEE Trans. Inf. Theory}, vol.~60, no.~6, pp.
  3238--3253, 2014.

\bibitem{satpathy2015gaussian}
S.~Satpathy and P.~Cuff, ``Gaussian secure source coding and {W}yner's common
  information,'' \emph{Proc. 2015 IEEE Int. Symp. Inform. Theory}, pp.
  116--120, June 2015.

\bibitem{yang2014wyner}
P.~Yang and B.~Chen, ``Wyner's common information in {G}aussian channels,'' in
  \emph{Proc. 2014 IEEE Int. Symp. Inform. Theory}.\hskip 1em plus 0.5em minus
  0.4em\relax IEEE, 2014, pp. 3112--3116.

\bibitem{li2016distributed}
C.~T. Li and A.~E. Gamal, ``Distributed simulation of continuous random
  variables,'' \emph{Proc. 2016 IEEE Int. Symp. Inform. Theory}, 2016.

\bibitem{arikan2009channel}
E.~Ar{\i}kan, ``Channel polarization: A method for constructing
  capacity-achieving codes for symmetric binary-input memoryless channels,''
  \emph{IEEE Trans. Inf. Theory}, vol.~55, no.~7, pp. 3051--3073, July 2009.

\bibitem{korada2009polar}
S.~B. Korada, ``Polar codes for channel and source coding,'' Ph.D.
  dissertation, Ecole Polytechnique F{\'e}d{\`e}rale de Lausanne, 2009.

\bibitem{PolarAsymOrig}
D.~Sutter, J.~Renes, F.~Dupuis, and R.~Renner, ``Achieving the capacity of any
  {DMC} using only polar codes,'' in \emph{Proc. 2012 IEEE Inform. Theory
  Workshop}, Sept. 2012, pp. 114--118.

\bibitem{aspolarcodes}
J.~Honda and H.~Yamamoto, ``Polar coding without alphabet extension for
  asymmetric models,'' \emph{IEEE Trans. Inf. Theory}, vol.~59, no.~12, pp.
  7829--7838, Dec. 2013.

\bibitem{PolarlatticeQZ}
\BIBentryALTinterwordspacing
L.~Liu and C.~Ling, ``Polar lattices for lossy compression,'' Jan. 2015.
  [Online]. Available: \url{http://arxiv.org/abs/1501.05683}
\BIBentrySTDinterwordspacing

\bibitem{GoelaCommonInfor}
N.~Goela, ``Polarized random variables: Maximal correlations and common
  information,'' in \emph{Proc. 2014 IEEE Int. Symp. Inform. Theory}, Honolulu,
  USA, June 2014, pp. 1643--1647.

\bibitem{cronie2010lossless}
H.~S. Cronie and S.~B. Korada, ``Lossless source coding with polar codes,'' in
  \emph{Proc. 2010 IEEE Int. Symp. Inform. Theory}, Austin, TX, June 2010, pp.
  904--908.

\bibitem{arikan2010source}
E.~Arikan, ``Source polarization,'' \emph{Proc. 2010 IEEE Int. Symp. Inform.
  Theory}, pp. 899--903, Austin, USA 2010.

\bibitem{polarlatticeJ}
\BIBentryALTinterwordspacing
Y.~Yan, L.~Liu, C.~Ling, and X.~Wu, ``Construction of capacity-achieving
  lattice codes: Polar lattices,'' Nov. 2014. [Online]. Available:
  \url{http://arxiv.org/abs/1411.0187}
\BIBentrySTDinterwordspacing

\bibitem{Ido}
I.~Tal and A.~Vardy, ``How to construct polar codes,'' \emph{IEEE Trans. Inf.
  Theory}, vol.~59, no.~10, pp. 6562--6582, Oct. 2013.

\bibitem{witsenhausen1976values}
H.~S. Witsenhausen, ``Values and bounds for the common information of two
  discrete random variables,'' \emph{SIAM J. Appl. Math.}, vol.~31, no.~2, pp.
  313--333, 1976.

\bibitem{nayak2010successive}
J.~Nayak, E.~Tuncel, D.~G{\"u}nd{\"u}z, and E.~Erkip, ``Successive refinement
  of vector sources under individual distortion criteria,'' \emph{IEEE Trans.
  Inf. Theory}, vol.~56, no.~4, pp. 1769--1781, 2010.

\bibitem{GeXulossyCI_conf}
G.~Xu, W.~Liu, and B.~Chen, ``Wyner{'}s common information for continuous
  random variables - {A} lossy source coding interpretation,'' in \emph{Proc.
  Annu. Conf. Inform. Sci. Syst.}\hskip 1em plus 0.5em minus 0.4em\relax IEEE,
  March 2011.

\bibitem{forney6}
G.~D. Forney~Jr., M.~Trott, and S.-Y. Chung, ``Sphere-bound-achieving coset
  codes and multilevel coset codes,'' \emph{IEEE Trans. Inf. Theory}, vol.~46,
  no.~3, pp. 820--850, May 2000.

\bibitem{LingBel13}
C.~Ling and J.-C. Belfiore, ``Achieving {AWGN} channel capacity with lattice
  {G}aussian coding,'' \emph{IEEE Trans. Inf. Theory}, vol.~60, no.~10, pp.
  5918--5929, Oct. 2014.

\bibitem{cong2}
C.~Ling, L.~Luzzi, J.-C. Belfiore, and D.~Stehle, ``Semantically secure lattice
  codes for the {G}aussian wiretap channel,'' \emph{IEEE Trans. Inf. Theory},
  vol.~60, no.~10, pp. 6399--6416, Oct. 2014.

\bibitem{multilevel}
U.~Wachsmann, R.~Fischer, and J.~Huber, ``Multilevel codes: Theoretical
  concepts and practical design rules,'' \emph{IEEE Trans. Inf. Theory},
  vol.~45, no.~5, pp. 1361--1391, July 1999.

\bibitem{GrayConditionalRD}
R.~M. Gray, \emph{Conditional rate-distortion theory}.\hskip 1em plus 0.5em
  minus 0.4em\relax Stanford, CA, Tech: Stanford Electronic labs., Oct 1972.

\end{thebibliography}

\end{document}